\documentclass[fleqn,usenatbib]{mnras}

\usepackage{newtxtext,newtxmath}

\usepackage[T1]{fontenc}
\usepackage{ae,aecompl}

\usepackage[pdftex]{graphicx}	
\usepackage{amsmath}	
\usepackage{amssymb}	

\usepackage{epsfig}
\usepackage{graphicx}
\usepackage{natbib}
\usepackage{lscape}
\usepackage{threeparttable} 
\usepackage{xspace}
\usepackage{color}
\usepackage{epstopdf}
\usepackage{gensymb}
\usepackage{multirow}
\usepackage{booktabs}
\usepackage{times}
\usepackage[utf8x]{inputenc}
\usepackage{array}

\begin{document}
\label{firstpage}
\pagerange{\pageref{firstpage}--\pageref{lastpage}}

\title[Internal shocks in 4U 0614+091]{Testing jet geometries and disk-jet coupling in the neutron star LMXB 4U 0614+091 with the internal shocks model}
\author[A. Marino]{\thanks{E-mail: alessio.marino@unipa.it}A. Marino$^{1,2,3}$,  J. Malzac$^3$, M. Del Santo$^2$, S. Migliari$^{4,5}$, R. Belmont $^6$, T. Di Salvo$^1$, 
\newauthor 
D. M. Russell$^7$, J. Lopez Miralles$^8$, M. Perucho$^8$, A. D'A\`i$^2$, R. Iaria$^1$, L. Burderi$^9$
\\
$^{1}$ Universit\'a degli Studi di Palermo, Dipartimento di Fisica e Chimica, via Archirafi 36,  I-90123 Palermo, Italy.\\
 $^2$INAF/IASF Palermo, via Ugo La Malfa 153, I-90146 - Palermo, Italy. \\
 $^3$IRAP, Universit\`e de Toulouse, CNRS, UPS, CNES, Toulouse, France.\\
 $^4$ XMM-Newton Science Operations Centre, ESAC/ESA, Camino Bajo del Castillo s/n, Urb. Villafranca del Castillo, 28691 Villanueva de la Ca\~nada, Madrid, Spain\\
 $^5$ Institute of Cosmos Sciences, University of Barcelona, Martí i Franquès 1, 08028 Barcelona, Spain\\
 $^6$ AIM, CEA, CNRS, Universit\`e Paris-Saclay, Universit\`e Paris Diderot, Sorbonne Paris Cit\`e, 91191 Gif-sur-Yvette, France\\
 $^7$ Center for Astro, Particle and Planetary Physics, New York University Abu Dhabi, PO Box 129188, Abu Dhabi, UAE\\
 $^8$ Departament d'Astronomía i Astrofísica, Universitat de València, Dr. Moliner 50, 46100, Burjassot (València), Spain\\
 $^9$ Universit\`a degli Studi di Cagliari, Dipartimento di Fisica, SP Monserrato-Sestu km 0.7, I-09042 Monserrato, Italy}
 
\date{Accepted XXX. Received YYY; in original form ZZZ}

\pubyear{2020}
\maketitle

\begin{abstract}
Multi-wavelength spectral energy distributions of Low Mass X-ray Binaries in the hard state are determined by the emission from a jet, for frequencies up to mid-infrared, and emission from the accretion flow in the optical to X-ray range. In the last years, the flat radio-to-mid-IR spectra of Black Hole (BH) X-ray binaries was described using the internal shocks model, which assumes that the fluctuations in the velocity of the ejecta along the jet are driven by the fluctuations in the accretion flow, described by the X-ray Power Density Spectrum (PDS). In this work we attempt to apply this model for the first time to a Neutron Star (NS) LMXB, i.e. 4U 0614+091. We used the multi-wavelength data set obtained in 2006, comprising data from radio to X-ray, and applied a model which includes an irradiated disc model for the accretion flow and an updated version of the internal shocks code for the ejection. The new version of the code allows to change the geometry of the jet for the case of non-conical jets. Only two alternative scenarios provide a satisfactory description of the data: using the X-ray PDS but in a non-conical geometry for the jet, or either using a conical geometry but with a "flicker-noise" PDS. Both scenarios would imply some differences with the results obtained with similar models on BH X-ray binaries, shedding light on the possibility that jets in NS and BH binaries might somehow have a different geometry or a different coupling with the accretion flow.
\end{abstract}


\section{Introduction}
The ejection of collimated outflows of matter in the form of jets is an ubiquitous phenomenon in astrophysics and it has been associated to a wide range of celestial objects, from Active Galactic Nuclei (AGNs) to Young Stellar Objects (YSOs) and stellar-mass X-ray Binaries (XRBs). Jets have been studied in detail for decades but many points about them, in particular concerning their launching mechanisms and their coupling with the accretion flow, are still debated \citep[see, for a review, ][]{Belloni2010}. \\
In XRBs, jets are usually not observed as extended structures, but their presence is witnessed by a radio-to-mid-IR flat spectrum \citep[in this case, jets are referenced to as 'compact', see, e.g.][]{Corbel2000, Fender2001, Corbel2002}. The spectrum of a compact jet is characterized by a jet break, corresponding to the base of the jet, where emission goes from optically thick to optically thin \citep[see, e.g. ][]{Gandhi2011,Russell2013,Koljonen2015} and at lower frequencies by a continuum which is the result of the superposition of the self-absorbed synchrotron spectra emitted from the different regions of the jet \citep{Condon1973, deBruyn1976, Marscher1977, Konigl1981, Ghisellini1985}. Moving along the jet, the magnetic field decays and the particles lose energy as the jet expands: the peaks of the local single synchrotron spectra are expected to decrease in both intensity and frequency, leading to an observed inverted radio spectrum \citep{Marscher1980}. However compact jet spectra are usually almost flat, and this has been explained in the past invoking the existence of some continuous energy replenishments mechanisms which would compensate the adiabatic losses due to the expansion of the jet \citep{Blandford1979}. Alternatively, sufficiently collimated jets would emit flat spectra even without taking into account dissipation mechanisms \citep{Kaiser2006}, but this would require such a fine tuning of the jet geometry that it is unlikely to explain the majority of the observed cases. \\
The nature of the dissipation mechanism is still unclear, but several models have been proposed in the past and involve, e.g., magnetic reconnection \citep{Sobacchi2020}, acceleration in relativistic shear flows \citep{Rieger2019} or Internal Shocks. The latters take into account the conversion from kinetic into internal energy which arises when two shells in the jet, ejected at different velocity, catch up and collide. Internal shocks models have been applied in the past to $\gamma$-ray bursts \citep{Rees1994,Daigne1998}, AGNs \citep{Rees1978,Spada2001,Boettcher2010} and BH XRBs \citep{Kaiser2000,Jamil2010,Malzac2013}. In the last decade, \citet{Malzac2013, Malzac2014} showed indeed that internal shocks can also explain the flat SEDs of XRBs if we assume that the fluctuations in the velocity of the ejecta are matched by the fluctuations in the accretion flow. Under this hypothesis, which underlies a profound accretion/ejection connection for jets in X-ray Binaries, 
one can use the observed X-ray Power Spectral Density as a proxy for the fluctuations of the accretion flow. The model, dubbed Internal Shocks model or ISHEM, has been successfully applied to a number of XRBs hosting black holes (BHs) as the primary star in the past \citep[][]{Drappeau2015,Baglio2018,Malzac2018,Peault2019,Bassi2020}, but never to a NS X-ray binary. \\
Jets are observed in X-ray binaries hosting NSs as well. They are very common in Low Mass X-ray Binaries (LMXBs) systems, which host NSs with usually weak ($B\lesssim10^9$ G) magnetic fields, but recently the phenomenon has been also associated to one highly magnetised NS in a High Mass X-ray Binary \citep{VanDenEijnden2018_jet}. 
Unfortunately, the wealth of studies of jet phenomenology in BH X-ray binaries is unmatched when it comes to NS X-ray binaries, mainly because they tend to have weaker radio emission (from hundreds to tens of $\mu$Jy), sometimes below the observational capabilities of the most sensitive interferometers on Earth. In addition NS LMXBs show faster state transition timescales \citep{MunozDarias2014,Marino2019b}, which make it harder to schedule coordinated space and ground observations. As mentioned before, jets in NSs are less radio-loud than in BHs, i.e. by a factor of $\sim$30 at similar X-ray flux levels \citep{FenderKuulkers2001, Migliari2003,Tudor2017,Gallo2018}. 
Furthermore, in BH binaries jet emission is always suppressed when the source is in the soft state, while jets are never entirely quenched in NS X-ray binaries \citep[see, e.g.][]{Migliari2004}, with just a few exceptions \citep{MillerJones2010, Gusinskaia2017}. Accretion-ejection coupling in binaries has been traditionally studied with radio--X-ray luminosity diagrams $L_{\rm R}$:$L_{\rm X}$ in both BH and NS XRBs and yet again a clearer picture seems to emerge for the former class of systems. In such diagrams\footnote{The largest available data base of X-ray/radio observations of XRBs is consultable at \url{ https://github.com/bersavosh/XRB-LrLx_pub}, while for the most recently published plot see \cite{Bassi2019}, fig. 7.}, BH XRBs tend to populate two branches $L_{\rm R}\propto L^{\beta}_{\rm X}$, i.e. with $\beta\sim 0.6$ for the "radio loud" systems and $\beta\sim 1.4$ for the "radio quiet" systems \citep[see, e.g.][]{Corbel2013}, and this behaviour has been proposed to originate in different physical properties in the accretion flow \citep{Coriat2011} or in the jet \citep{Espinasse2018} over the two branches. A similar dichotomy can be found in NS XRBs, but the distribution appears more scattered and harder to interpret \citep{Tetarenko2016}. Alternatively, \cite{Gallo2018} proposed a single-track population for both BH and NS XRBs but with different values of $\beta$, i.e. the formers with $\beta\sim$0.7 and the latters with $\beta\sim$0.4. 
It is also noteworthy that even the jet launching mechanism could not be the same in BH and NS LMXBs, since of the two traditionally proposed jet launching mechanisms, i.e. \cite{BlandfordZnajek1977} and \cite{BlandfordPayne1982}, only the second could be at work in both classes \citep[but see also ][ for a discussion on the possibility of having spin-powered jets also on NS LMXBs.]{Migliari2011} In summary, the emerging picture seems to imply that the nature of the compact object (and, as in the case of a NS, the presence of a magnetic field) could play some role in determining the characteristics and the origin of the jet. \\
\subsection{4U 0614+091}
\label{ss:source}
Discovered by the Uhuru survey in the 70s \citep{Forman1978}, 4U 0614+091 was later identified as a Low Mass X-ray binary hosting a NS by the detection of type-I X-ray bursts \citep[]{Swank1978}. From the study of the bursts a measure of the distance was obtained, i.e. around 3.2 kpc with a 15\% uncertainty \citep{Kuulkers2010}.  
Due to its short orbital period of around 50 mins \citep[]{Shahbaz2008,Baglio2014}, the system has been classified as an ultra compact X-ray binary or UCXB, implying a likely degenerate-helium dwarf or white dwarf nature for the companion star \citep[see, e.g.][ for an extensive discussion on the companion star nature]{Kuulkers2010}. 
The source is classified as a persistent atoll source\footnote{Low luminosity NS LMXBs which exhibit mainly two spectral states, one hard, dubbed "island" state, and one soft, dubbed "banana state" \citep[][]{Hasinger1989}.} and is expected to spend in the hard ("island") spectral state almost 90\% of its time \citep{Vanstraaten2000}, with a constantly high X-ray spectral variability level (higher than 5\%) and only episodical transitions to softer states \citep{MunozDarias2014}. In the past, a few authors gave evidence for the presence of a so-called hard tail in the spectrum, reaching energies beyond 100 keV, which was modeled with non-thermal models \citep{Piraino1999,Migliari2010} or thermal Comptonization from a very hot corona \citep{Ford1997,Piraino1999,Fiocchi2008}. A reflection component has been commonly used to describe the X-ray spectral emission too, although the Fe K line was usually found to be absent or weak. The apparent absence (or weakness) of this feature may be related to an underabundance of Fe \citep[][]{Madej2014, Ludlam2019} in the secondary star with respect to solar abundances and it is compatible with the hypothesis of an out-of-main-sequence companion. \\
The flat radio to mid-IR spectrum, reported by \cite{Migliari2010} in the first complete multi-wavelengths spectral study of the source, witnesses the presence of a compact jet in the system, as confirmed by the polarimetric study by \cite{Baglio2014}. \\
In this paper we report on the application of the Internal Shocks model to the Spectral Energy Density of 4U 0614+091, the first attempt ever to describe the entire SED of a NS LMXB with a model including both the jet and the accretion flow emission. Furthermore, ISHEM has been used so far only with systems hosting a black hole and never for a NS LMXB, as in this work. 
\section{Data} \label{sec:data}
The ISHEM model depends on three essential ingredients: (1) a multi-wavelength SED, (2) a PDS which is used as tracer of the accretion flow variability and (3) a synthetic spectrum, simulated on the basis of the PDS to be compared with the real SED. In the previous applications of ISHEM \citep[see][]{Drappeau2015,Malzac2018, Peault2019, Bassi2020}, it was proven that the X-ray PDS quasi-simultaneous to the SED can be satisfactorily used as ingredient (2). The methods to obtain these three ingredients are described in the following sections. In particular in this Section we describe the data set and the timing analysis while the model used, the main parameters adopted and the spectral fitting procedure will be described in Section \ref{sec:ishem}. \\
This work takes advantage of the multi-wavelength observational campaign performed on the source within 5 days, between October 30 and November 4 2006, from radio to X-ray. For the radio-to-IR realm, we used radio observations collected by the Karl G. Jansky Very Large Array (\emph{VLA}), mid-IR/IR observations taken by the Infrared Array Camera (IRAC) onboard \emph{Spitzer Space Telescope}. We used near-IR/optical data by the ground based Small and Moderate Aperture Research Telescope System (\emph{SMARTS}), while the optical/UV observations were taken with UVOT onboard the Neil Gehrels Swift Observatory (\emph{Swift} in the following). Finally X-ray data were obtained with the Proportional Counter Array (PCA) and the High Energy X-ray Timing Experiment (HEXTE) onboard the Rossi X-ray Timing Explorer (\emph{RXTE}) and XRT onboard \emph{Swift}. This vast data set was already used for a comprehensive spectral analysis by \cite{Migliari2010}. We refer to this paper for the details concerning the data reduction and analysis performed on the observations, except for the \emph{Swift}/UVOT data, which were re-extracted in this work.  
\subsection{\emph{SMARTS}-UVOT data treatment}
\begin{figure}
\centering
\includegraphics[scale=0.40]{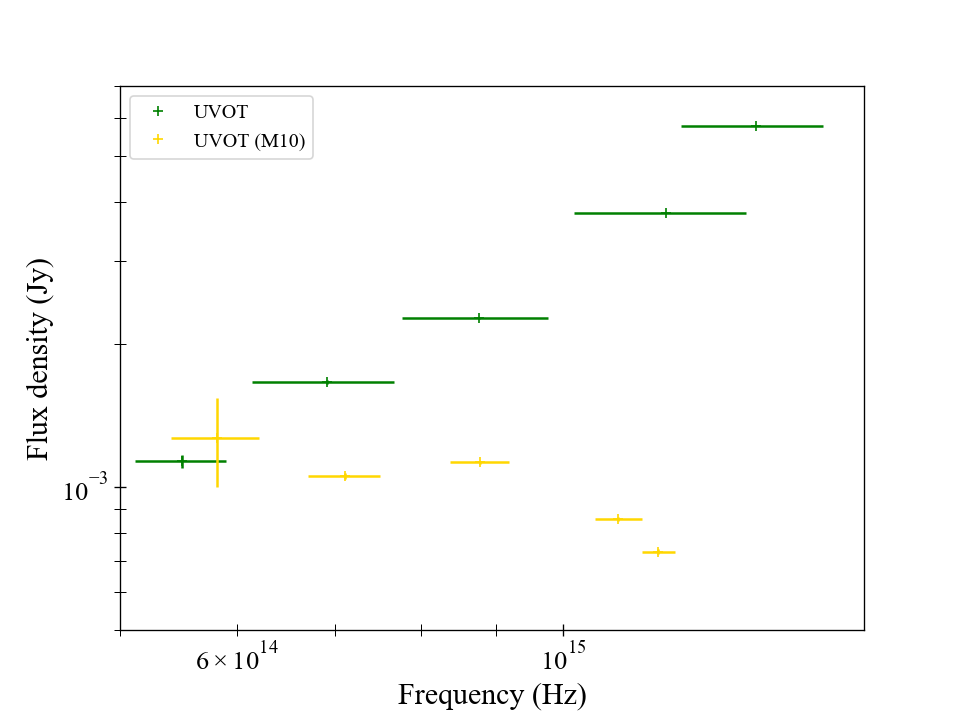}
\caption{Comparison between the unabsorbed UVOT data in \citet{Migliari2010} (yellow points) and the unabsorbed UVOT data used in this paper (green points). In particular, for the new UVOT data set we corrected for an absorption coefficient of A$_{\rm V} \sim$1.5, found using the model \textsc{redden} on \textsc{Xspec} (see subsection \ref{ss:diskemi}-\ref{ss:global} for further details).}
\label{fig:uvot}
\end{figure}

As pointed out by \cite{Migliari2010}, the data in the optical-ultraviolet (covered by \emph{SMARTS}-UVOT) region of the electromagnetic spectrum show indeed an unexpected shape, which can not be ascribed to irradiation of the outer disk or to the blackbody emission from the \citep[likely very faint,][]{Nelemans2004, Shahbaz2008} companion star. In order to check if a different treatment of the data might improve their results, we re-analysed UVOT data (ObsID 00030812001) with HEASOFT v. 6.26 following the standard procedure\footnote{reported in \url{https://www.swift.ac.uk/analysis/uvot}} and we used the calibration files updated to the latest available version (CALDB 2017-09-22). This observation was carried out using all the six UVOT filters. By using the task \textsc{uvotdetect}, we clearly detected 4U 0614+091 in each image. We defined a source region with a 5 arcsec radius and 
several different background regions for each filter around the source.
Finally, the photometry of 4U 0614+091 has been performed with the task 
\textsc{uvotsource}. \\
The de-reddening applied by \cite{Migliari2010} on the \emph{SMARTS} data was removed by using equation (1) and the values reported in table 3 (for the bands V, I and J) of \cite{Cardelli1989} and considering A$_{\rm V}$=2, as reported in \cite{Migliari2010} in order to get A($\lambda$), i.e. the absorption at wavelength $\lambda$. Considering then $F_\lambda = F_{\lambda,0}\times e^{-A(\lambda)/1.086}$, with $F_\lambda$ and $F_{\lambda,0}$ the absorbed and unabsorbed fluxes at wavelength $\lambda$ respectively, we obtained the required values for the uncorrected \emph{SMARTS} data. The new UVOT data and the uncorrected \emph{SMARTS} data have been then de-reddened by us via \textsc{Xspec}, using a proper model (see Subsection \ref{ss:diskemi}). \\
With the new treatment of the optical-UV data the odd IR-UV spectral shape reported by \cite{Migliari2010} has now disappeared. In Figure \ref{fig:uvot} we compare the "old" and the new data sets de-reddened, in order to investigate the nature of the previously reported tricky result. We therefore retain this discrepancy arises from the different extraction methods, in particular on the choice and sizes of source and background regions used for the photometric measures.

\subsection{X-ray Timing Analysis}\label{ss:timing}
\begin{figure}
\centering
\includegraphics[scale=0.37]{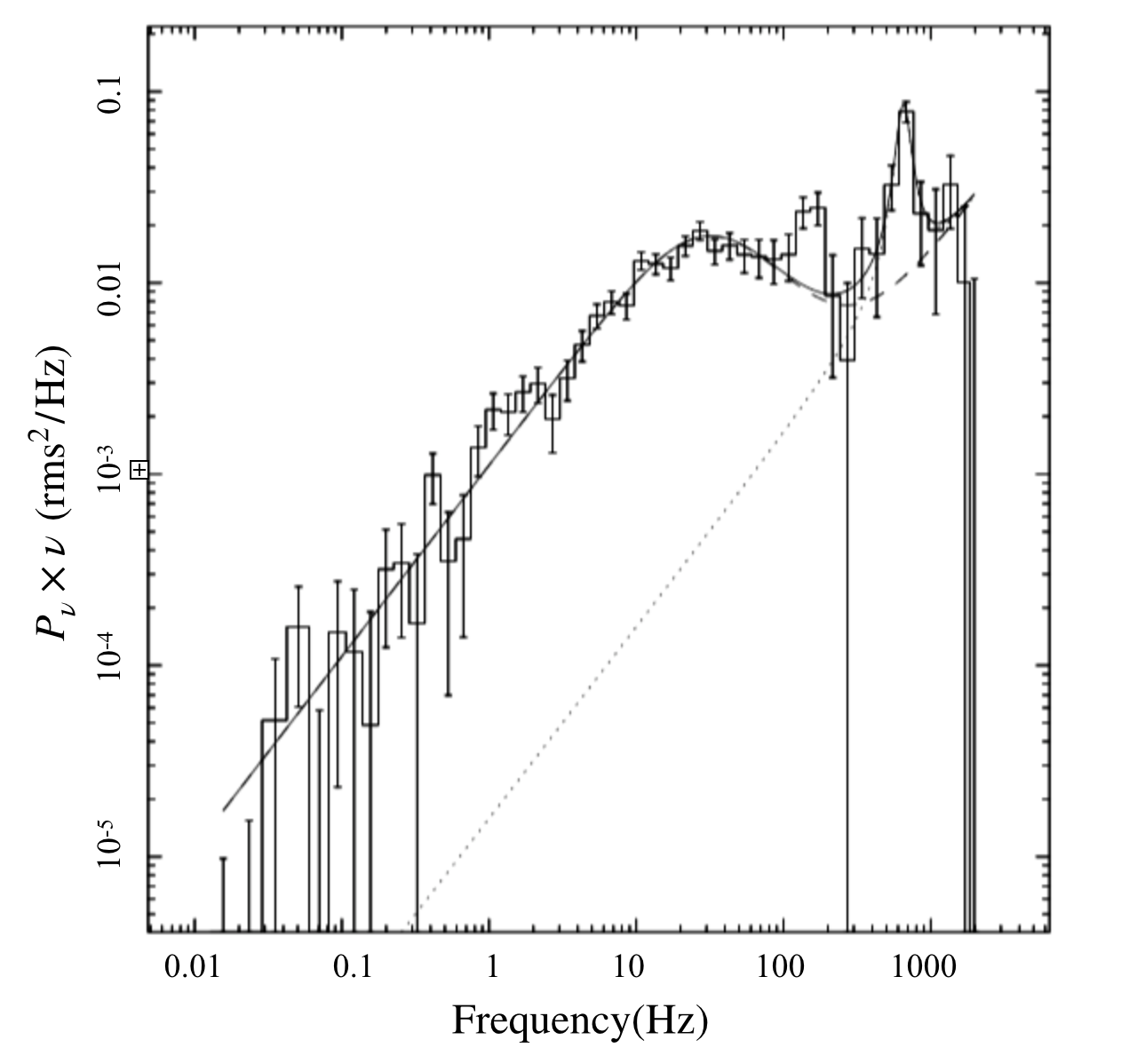}
\caption{PDS in the normalized power ($P_{\nu}$) times frequency ($\nu$) representation, with the best fit multi-Lorentzian model. }
\label{fig:pds}
\end{figure}

For the timing analysis of the observation 92411-01-06-07 (30 October 2006), we used RXTE Proportional Counting Array (PCA) data in the \textit{event} mode configuration with a time resolution of $\sim$125 $\mu$s, allowing to obtain PDS up to a Nyquist frequency of 4096 Hz. We averaged multiple PDS data calculated over 128 seconds subintervals covering a total data set of 2048 seconds, using Fast Fourier Transform (FFT) techniques. No deadtime corrections nor background subtraction were performed before creating the PDS. We subtracted the Poisson noise power, derived from the PDS in the frequency range 1536 and 2048 Hz, following \cite{Zhang1995}. Figure \ref{fig:pds} shows the traditional $\nu$, $P_{\nu}$ representation where we applied the Leahy normalisation \citep{Leahy1983} before converting the PDS to squared fractional rms. {The resulting PDS was fitted with a model consisting of the sum of two Lorentzians, one broad Lorentzian to fit the low frequency noise and one narrow to fit the QPO in the range $\sim$500-700 Hz.  The best-fit parameters found were used as input for ISHEM and are listed in Table \ref{tab:pds}.} 

\begin{table}
\centering
\caption{Fit results of the PDS described with a sum of two Lorentzians, each of them given by $P(\nu)=r^2/\pi\left[\Delta^2+(\nu-\nu_0)^2\right]$, with $r$ the integrated \emph{rms} over the full range of frequencies $-\infty$ to $+\infty$, $\Delta$ the Full Width Half Maximum of the Lorentzian and $\nu_0$ its central frequency. Values in round parentheses were kept frozen during the fit. Quoted errors reflect 68\% confidence levels.}
\begin{tabular}{ l  l l l}
\hline 
\hline
Lorentzian & \multirow{2}{*}{$\nu_0$} & \multirow{2}{*}{$\Delta$} & \multirow{2}{*}{$r$} \\
Component \\
\hline
1 & (0) & $62.0\pm 5.0$ & $0.106^{+0.003}_{-0.006}$\\
2 &  $650^{+30}_{-24}$ & <250 & $0.030^{+0.010}_{-0.008}$ \\
\hline
& \multicolumn{3}{r}{$\bf{\chi^2_{\nu}}(d.o.f.)$ = 1.49(126)}\\
\hline
\hline
\end{tabular}
\label{tab:pds}
\end{table}

\section{The jet model} \label{sec:ishem}

\subsection{The spectral shape}\label{ss:shape}
The \textsc{ISHEM} model aims to describe the spectral energy distribution of jets based on how the energy is dissipated along the jet. \\ As mentioned before, the emission of jets is ascribed to the superposition of the self-absorbed synchrotron spectra emitted locally from the different regions of the jet, peaking at decreasing energies as we move away from the base of the jet. However, the almost "flat" final spectrum requires that in roughly each region the energy lost in the expansion of the jet is somehow gained back by the particles. If we imagine the jet as the result of the periodic ejection of discretized shells of matter and if these shells are ejected with variable velocity or Lorentz factor $\Gamma$, then we expect a fraction of the energy of these shells to be released for each collision between ejecta travelling with different $\Gamma$. Internal shocks turn out then to be a viable mechanism to replenish the energy lost and thereby to flatten the spectra. However, shocks have to occur homogeneously all over the jet axis. The required dissipation pattern is ultimately determined by two factors: how fast the flow expand and the particles lose energy, which is in turn determined by the geometry of the jet (1) and how the velocity of the ejected shells fluctuates over time (2). Indeed, a fast variability, i.e. over short time scales, mainly produces collisions close to the base of the jet while on the other hand shells subject to slow variability would tend to produce shocks at higher distances. 
In \textsc{ISHEM} these two factors are regulated by a geometry parameter $\zeta$ and the input Power Density Spectrum (PDS) $P(\nu)$ (with $\nu$ frequency) of the Lorentz factor $\Gamma$ fluctuations. A third important ingredient is $p$, which determines the slope $\alpha=(p-1)/2$ of the spectrum at high energies, in the optically thin part of the jet spectrum. The final SED shape is determined by a combination of these three ingredients. In the following we will give more details on the effects of the two aforementioned factors.
\subparagraph{PDS}: In order to have a homogeneous dissipation pattern, variability in the ejecta must be almost homogeneously distributed over a large range of time-scales. For example, in a conical geometry, the exact compensation of the energy losses requires that the PDS of the jet Lorentz factor fluctuations corresponds  to a “flicker noise”, i.e.  $P(f) \propto 1/f $  over a broad range of Fourier frequencies \citep{Malzac2013}. Such flicker noises  occur if variability in the jet bulk Lorentz factor $\Gamma$ is dominated by the so-called "flicker noise" \citep[see, e.g.][]{Press1978}, which occurs in processes of different nature, e.g. biological, economic and physical, but also in astrophysics, especially in the X-ray variability of X-ray binaries \citep{Gilfanov2010}.
Interestingly, the Fourier PDS extracted from the X-ray light curves of BH XRBs in the hard spectral state is, at low frequencies, very similar to a flicker-noise PDS \citep[see ][ and references therein]{Malzac2014}. Furthermore, in a disk-jet coupling scenario, shells are ejected from the disk and it is reasonable to expect that the fluctuations in the accretion flow might be transmitted to the ejecta. According to these clues, the X-ray PDS can serve as the required $P(\nu)$ in \textsc{ISHEM}, as in e.g. \cite{Drappeau2015}. \\

\begin{figure}
\centering
\includegraphics[scale=0.27]{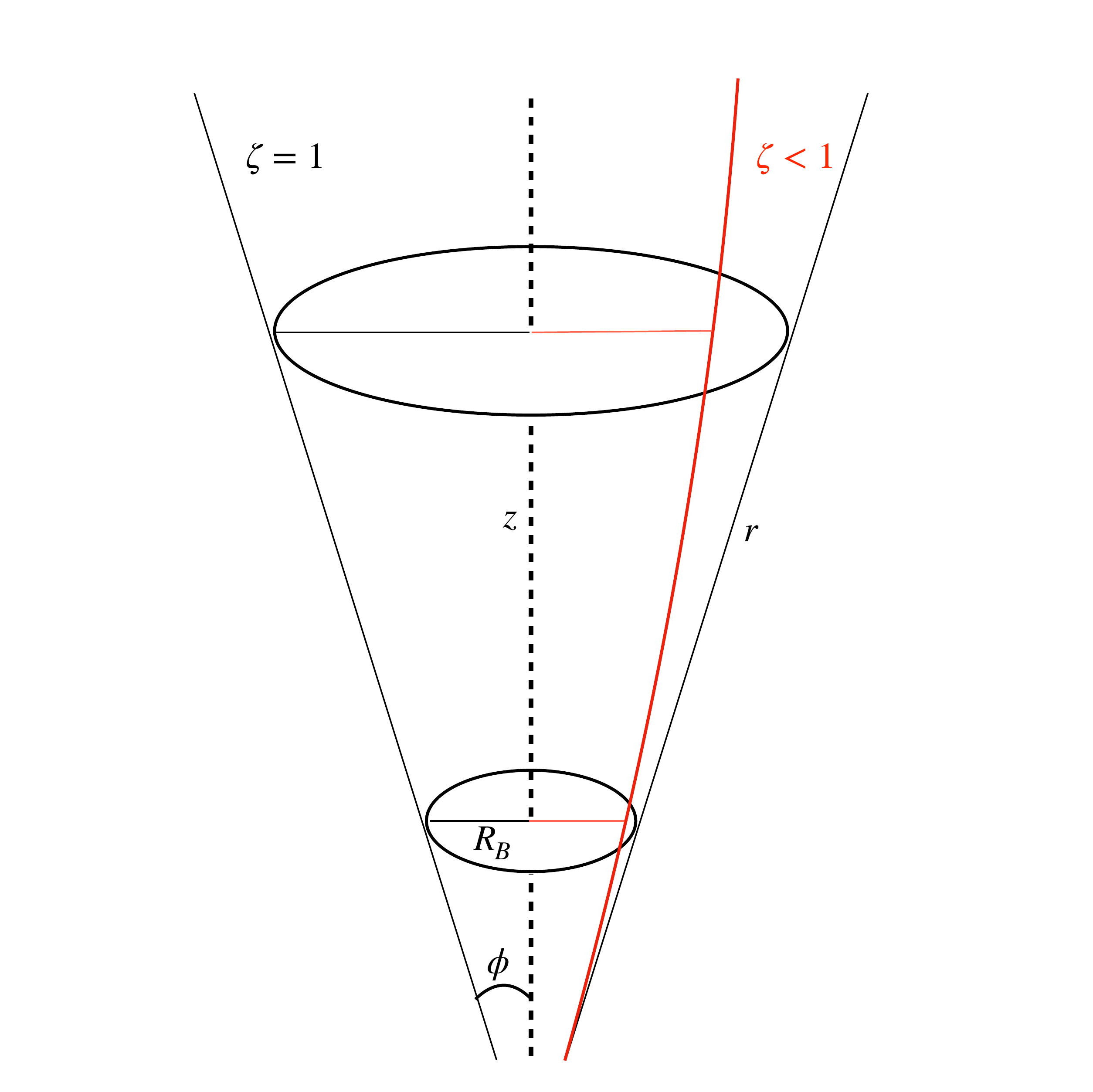}
\caption{Schematic view of a jet in both conical (black) and non-conical geometry (red).}
\label{fig:jet}
\end{figure}

\begin{figure}
\centering
\includegraphics[scale=0.37]{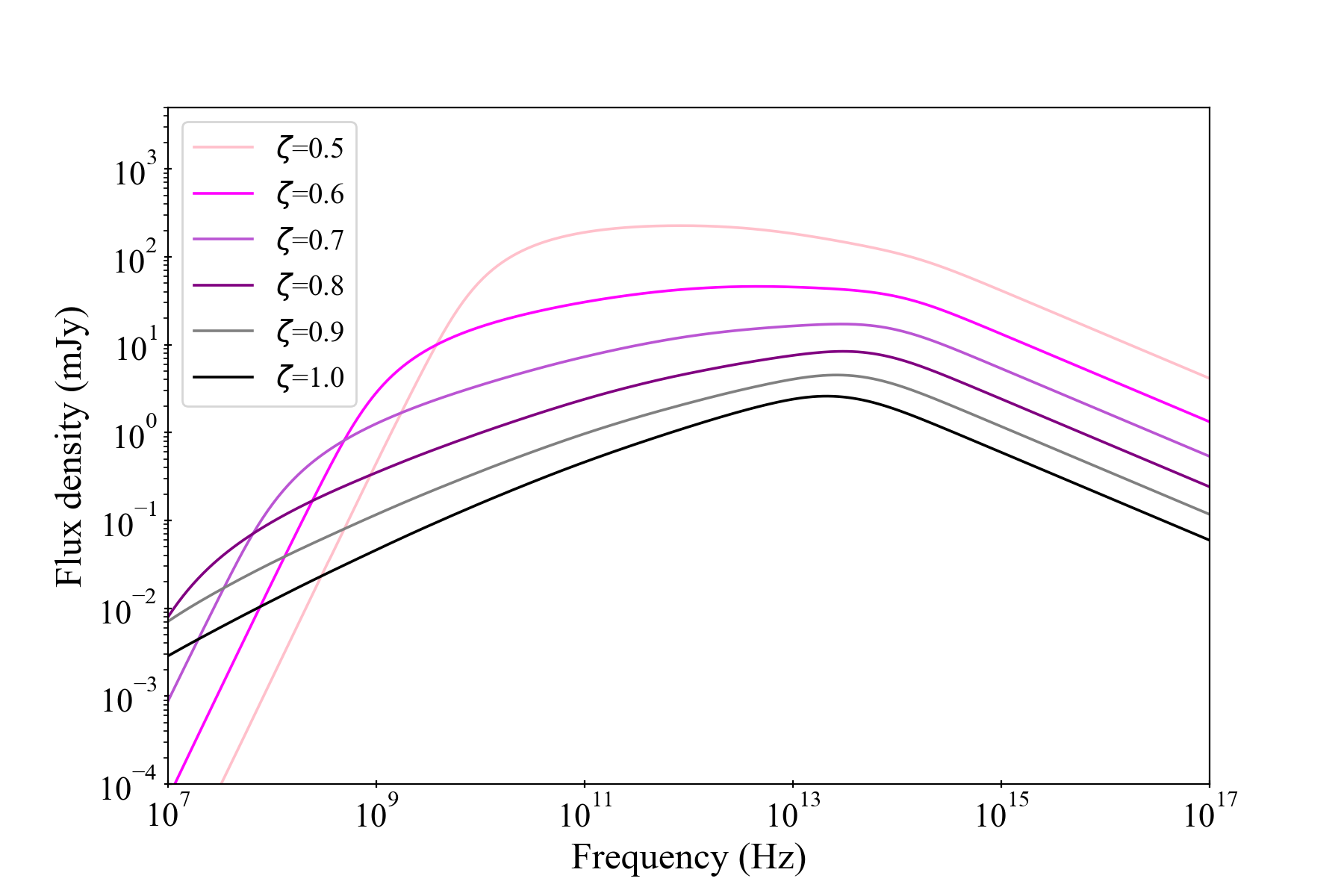}
\caption{Simulated SEDS with ISHEM using several values of $\zeta$, from 1.0 to 0.5, without rescaling or shift factors applied, showing the impact of geometry on the flux and the shape of the emitted jet spectrum. In order to simulate these spectra, the X-ray PDS was used as input.}
\label{fig:geometriesnonorm}
\end{figure}

\begin{figure*}
\centering
\includegraphics[scale=0.37]{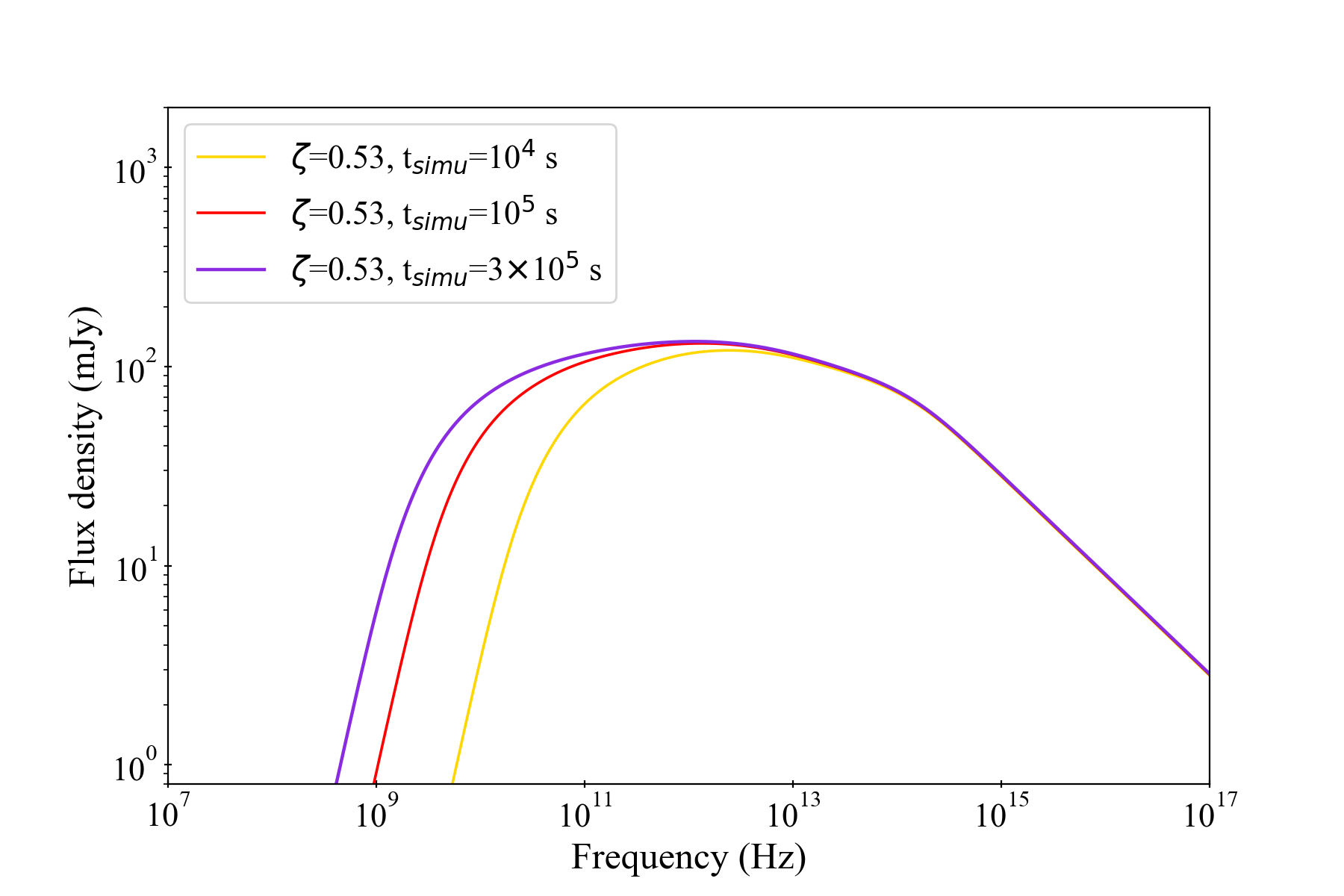}
\includegraphics[scale=0.37]{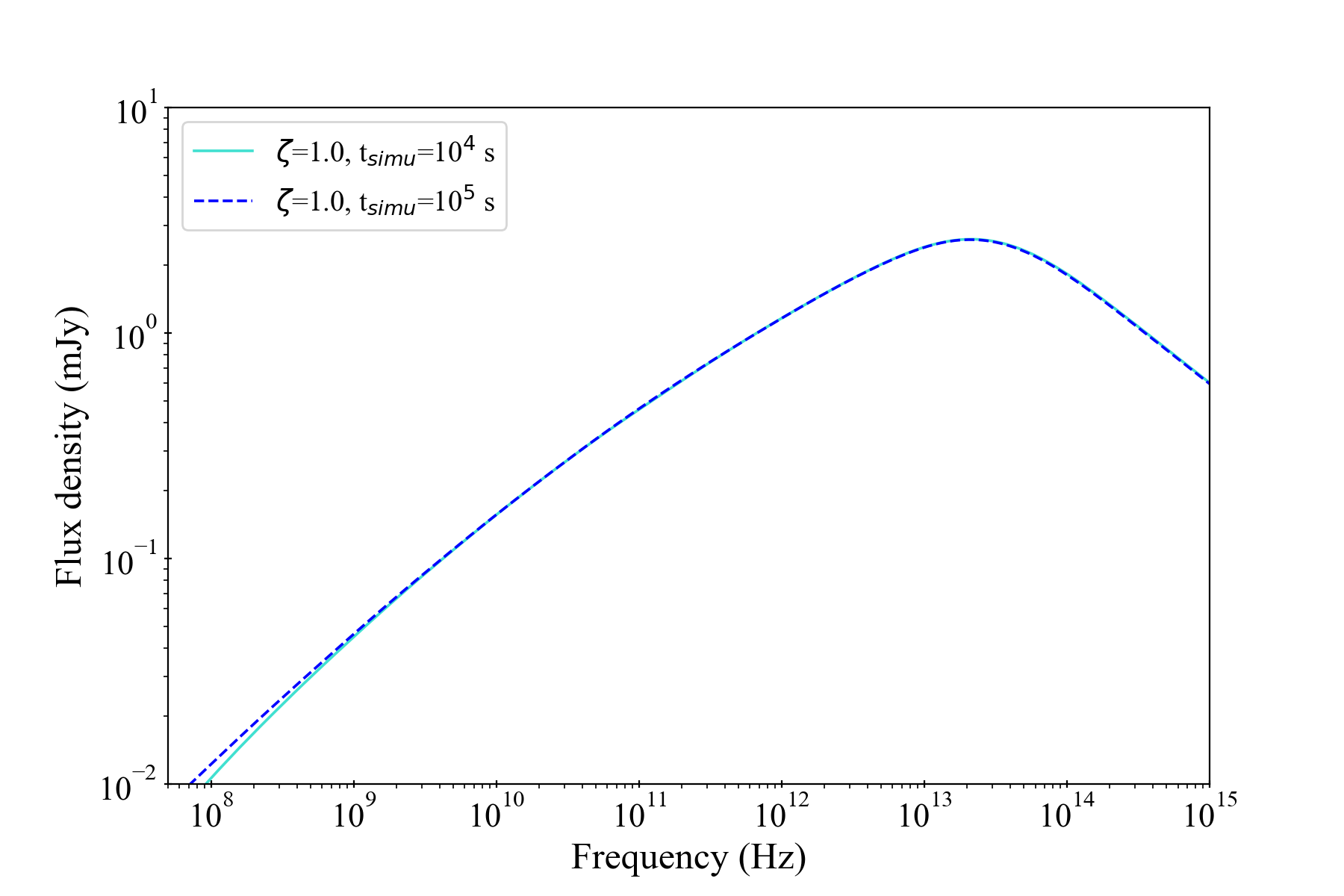}
\caption{Simulated SEDS with ISHEM using several values for $t_{\rm simu}$ in both non-conical (\emph{left}) and conical (\emph{right}) geometry, showing how for strongly non-conical jets the low frequency turnover is dependent on the choice for $t_{\rm simu}$. For both plots, $t_{\rm simu}$ was fixed to 10$^5$ s.}
\label{fig:tsimu}
\end{figure*}

\begin{figure}
\centering
\includegraphics[scale=0.37]{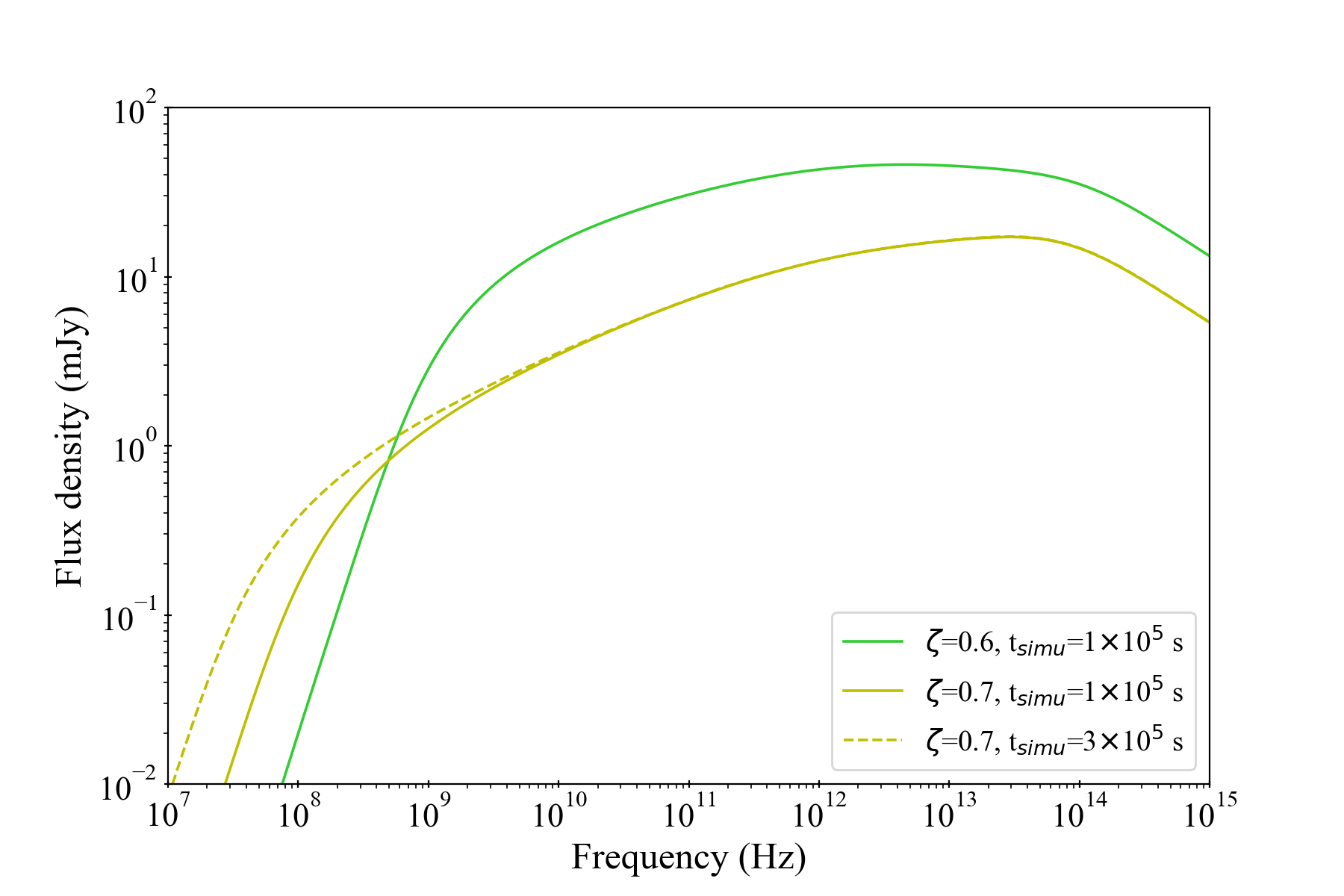}
\caption{Simulated SEDS with ISHEM using several values for $t_{\rm simu}$ and $\zeta$ in order to show the separate effect of reducing $\zeta$ and increasing $t_{\rm simu}$.}
\label{fig:tsimu_2}
\end{figure}

\subparagraph{Geometry}: Usually jets are assumed to be conical, i.e. the radius of the jet $r$ at a height $z$ follows a simple linear relation. 
However internal or external agents, as e.g. a toroidal component of the magnetic field in the jet or the pressure exerted by the interstellar medium (see Section \ref{sec:disc} for a discussion on the collimation agents) may collimate the jet and change its shape from conical to parabolic. More specifically, we can describe the geometry of the jet with a parameter $\zeta$ such as: $r \propto z^{\zeta}$, where $\zeta=1$ for the "standard" conical geometry, while $\zeta<1$ holds for a parabolic jet (see Figure \ref{fig:jet}) and the latter is likely a physically more realistic description of the jet structure. The case of an "overpressured jet" with $\zeta>1$ is plausible, but it would be a highly unstable structure which will tend to evolve spontaneously to a situation where $\zeta\leq1$ \citep{Kaiser2006} \footnote{This phenomenon has been predicted for parsec and kpc-scale AGN jets, where the overexpansion triggers pinching and periodic recollimation shocks, which force $\zeta$ to become lower than 1 \citep[see, e.g.][ and references therein.]{Perucho2007,Godfrey2012,Fromm2016}}. In the previous applications of \textsc{ISHEM}, the geometry was assumed to be conical and $\zeta$ was fixed to 1 by default. This is the first time that the dependency of the results on $\zeta$ is tested. In order to correctly take into account a non-conical geometry, the code used in this work has been updated with respect to the previous versions used by, e.g., \cite{Drappeau2015, Peault2019}. The new version of the model also includes some improvement of the treatment of radiation transfer, detailed in Appendix \ref{sec:model}. Some examples of simulations illustrating the effects of a non-conical geometry are shown in Fig. \ref{fig:geometriesnonorm}. As shown in the figure, reducing $\zeta$ allows for a more collimated jet, where the energy losses are reduced and the spectrum is naturally flatter. Furthermore, for a fixed length of the jet, the range of wavelengths emitted by the different regions will shrink as a result of the contained energy losses. This leads to the appearence of a low frequency turn-over which marks a transition from the flat partially absorbed region of the SED to  optically thick emission $F_{\nu}\propto \nu^{5/2}$ at lower frequencies. This optically thick emission arises from the terminal part of the jet at the largest scale. As can be seen in Fig.~\ref{fig:geometriesnonorm}, the low frequency termination turn-over gradually moves toward higher frequencies at lower $\zeta$ and might be observable if the jets are strongly confined. 
Lowering $\zeta$ results also in an increase of the overall flux emitted, as both the amount of energy lost and the frequency range over which jet power is distributed diminishes. 
The frequency of the termination break in the SED of the source also depends on the size of the jet, which, in our model corresponds to the distance to which the shells have been able to propagate during the time of the {\sc ishem} simulation i.e. $\sim c t_{\rm simu}\simeq 3\times 10^{15}$ cm in the spectra shown in Fig.~\ref{fig:geometriesnonorm}\footnote{Note that in the real world, the extension of the jet depends not only on the time since the ejection started, but also on the interaction of the jets with their ambient medium at large scales which is not modelled here}. 
As shown in Fig.~\ref{fig:tsimu}, in the case of strongly parabolic jets, increasing the simulation time pushes the spectral turnover towards lower frequencies, without affecting the shape of the spectrum above the turn-over frequency (see Fig. \ref{fig:tsimu_2} for a distinction between the effect of reducing $\zeta$ and increasing $t_{\rm simu}$). In contrast, in the conical jet model the SED is barely affected by the duration of the simulation. \\ 

\subsection{ISHEM parameters}\label{ss:param}
A wide set of other physical parameters is taken into account by the code as well, but they can only shift in frequency or scale the normalisation of the whole SED without modifying its shape. These parameters describe the system, the jet and the distribution of the radiating particles. The main parameters are: the distance ($D$) to the source, the inclination of the jet axis with respect to the line of sight ($\theta$), the mass of the compact object ($M_{\rm NS}$ in this case), the jet power ($P_{\rm J}$), the jet opening angle ($\phi$), the radius at the base of the jet ($R_{\rm b}$), the average Lorentz factor ($\Gamma_{\rm av}$) of the ejecta, the volume filling factor $f_{\rm V}$ \citep{Malzac2013}, the maximum/minimum energy limits ($\gamma_{\rm max}$,$\gamma_{\rm min}$) of the electron distribution. As mentioned in Subsection \ref{ss:source}, the distance of the source is well constrained to be around 3.2 kpc \citep{Kuulkers2010}, while no constraints have been ever reported to our knowledge on the mass of the NS, which in the following will be fixed to 1.5 M$_{\rm \odot}$\footnote{Which is close to the peak for recycled NSs in the expected NSs mass distribution \citep{Ozel2012}}, or to the inclination of the jet $\theta$, or to the inclination $i$ of the system itself. We therefore tentatively fix $\theta$ to 60$^\circ$. 
The jet power $P_{\rm J}$ is not known, so we fixed it to be of the same order of magnitude of the X-ray luminosity of the source, i.e. 0.01 $L_{\rm Edd}$ \citep{Migliari2010}. The effects of $P_{\rm J}$ and $\theta$ will be explored in more detail in Section \ref{sec:disc}. \\
We chose $f_{\rm V}=0.7$ \citep{Malzac2014} and $R_{\rm b}$ equal to 10 R$_{\rm G}$, which is plausible for X-ray binaries, although the impact of these parameters on the overall results is negligible. For $\Gamma_{\rm av}$, which for X-ray binaries is expected to vary between 1 and 10 \citep[see, e.g][]{Casella2010, Saikia2019}, we started with a value of 2 \citep{Gallo2003,Heinz2004}. We adopted a value of $\phi$ of 2$^\circ$, as opening angles are expected to be $\leq 10^\circ$ \citep{MillerJones2006} and a value of 2$^\circ$ has been used in the past\footnote{However, it was recently suggested that jet opening angles in XRBs could be even smaller, below 1$^\circ$ \citep{Zdziarski2016}.} \citep[see, e.g ][]{Stirling2001}. For the lower and upper limits of the electron distribution, we started with some standard values for X-ray Binaries, i.e. 10 and 10$^6$ respectively \citep{Gandhi2011,Malzac2014, Drappeau2015}. A study of how the resulting simulated SED depends on these parameters is presented in \cite{Peault2019}, figure 2. \\
As already discussed in section~\ref{ss:shape}, in the case of non-conical jets ($\zeta < 1$), the choice for $t_{\rm simu}$ becomes crucial because it determines both the size of the jet and the location of the low-frequency jet termination turn-over. We choose to set this parameter to $10^5$ s, which corresponds to a final jet extension of $\sim3\times 10^{15}$ cm. As the observed radio spectrum in 4U 0614+091 is rather flat reproducing the data with a strongly non-conical model will require to have the spectral turn-over well below 10 GHz. So we want the jet to be as large as possibly allowed by the observational constraints. Regarding 4U 0614+091 the constraints on the jet extension are very poor. Namely, the jets from 4U 0614+091 should not be  significantly bigger than $\sim 10^{17}$ cm, otherwise they would have been resolved with the VLA. In general, however, the observations of compact jets suggests much smaller dimensions for the radio emitting region. In the case of the resolved jet of Cyg X-1 the extension of the radio jet at 8.4 GHz indicates scales of the order of $10^{14}-10^{15}$ cm \citep[see, e.g ][]{Stirling2001}. Our choice for $t_{\rm simu}$ is therefore the most favourable for non-conical jet models while being still roughly compatible with the expected scale of the radio jet.  

A summary of the parameters used in the simulation is reported in Table \ref{tab:param}. 

\begin{table} \label{tab:ishem}
\centering
\caption{Parameters used in the \textsc{ISHEM} code which were kept fixed in all the simulations run in this paper. $a$: Simulation running time; $b$: Initial radius of the ejecta, imposed of approximately the same order of magnitude of the inner radius of the accretion disk; $c$: effective adiabatic index of the flow, \citep{Malzac2013}. }
\begin{tabular}{l  l  }
\hline 
\hline
\multicolumn{2}{l}{\textbf{Simulation parameters}} \\
\hline
 M$_{\rm NS}$ (M$_\odot$) & 1.5 \\
 t$_{\rm simu}^a$ (s) & 10$^5$ \\
 r$_{\rm dyn}^b$ (R$_{\rm G}$) & 10\\
 $\phi$ ($^\circ$) & 2.0\\
 f$_{\rm vol}$ & 0.7\\
 $\gamma_a^d$ & 4/3\\
 $\gamma_{\rm average}$ & 2.0\\
 Ejecta scheme & constant shell mass\\
 $P_{\rm jet}$ ($L_{\rm Edd}$) & 0.01 \\
 $\gamma_{\rm min}$ & 10\\
 $\gamma_{\rm max}$ & 10$^6$\\
 $\theta$ ($^\circ$) & 60\\
\hline
\hline
\end{tabular}
\label{tab:param}
\end{table}

\subsection{Simulations}
In the previous sections, we gave details on the data set and on the model. In order to test if data and model are compatible it is necessary first of all to compute a simulated, \emph{synthetic} SED using \textsc{ISHEM}. The code simulates over a fixed simulation running time $t_{\rm simu}$ the ejection of shells with velocity variable according to the input PDS in an environment which is set-up by the choice of $p$, $\zeta$ and the parameters described in Subsection \ref{ss:param}. The simulated SED is produced to build a local model on \textsc{Xspec} (v. 12.10.1f) called \textsc{ish} used to fit the data. The model is characterized by basically two parameters, i.e. a re-normalization parameter and a shift parameter, which allow to rescale or shift in frequency the synthetic SED but not to change its shape, determined by the parameters set-up in \textsc{ISHEM}.
The model \textsc{ish} is therefore used to fit the data. In the case of a poor fit, a different combination of PDS, $\zeta$ and $p$ needs to be used in order to change the spectral shape. When a good fit is found, the best-fit scaling and shift parameters can be used to improve the original set of parameters in Subsection \ref{ss:param}. The shift parameter scales as the frequency break $\nu_{\rm b}$ and the renormalization parameter scales as the flux at this frequency $F_{\nu_{\rm b}}$. The following system of relation holds for these parameters:
\begin{equation} \label{eq:fluxbr}
{F}_{\nu_{\rm b}} \propto \frac{R_{b}^{\zeta-1}}{\tan^\zeta{\phi}}\,
\frac{\delta^\frac{3p+7}{p+4}}{D^2_{\rm kpc}}\,\frac{ \sin{\theta}^\frac{p-1}{p+4}\quad i_\gamma^\frac{5}{p+4}\quad P_J^\frac{2p+13}{2p+8}}
{ \left[(\Gamma_{\rm av}+1)\Gamma_{\rm av}\beta\right]^{\frac{2p+13}{2p+8}+\zeta-1}}
\end{equation}
\begin{equation} \label{eq:break}
\nu_{\rm b}\propto \frac{R_{b}^{\zeta-1}}{\tan^\zeta{\phi}}\,
\delta^\frac{p+2}{p+4} \,
\frac{ 
\sin{\theta}^\frac{-2}{p+4} \quad i_\gamma^\frac{2}{p+4} \quad P_J^\frac{p+6}{2p+8}}
{\left[(\Gamma_{\rm av}+1)\Gamma_{\rm av}\beta\right]^{\frac{3p+14}{2p+8}+\zeta-1}}
\end{equation}

where $\beta=\sqrt{1-\Gamma^{-2}_{\rm av}}$, $\delta=\left[\Gamma_{\rm av}(1-\beta \cos{i})\right]^{-1}$ and $i_\gamma^{-1}=\int_{\gamma_{\rm min}}^{\gamma_{\rm max}} \gamma^{-p}(\gamma-1)d\gamma$. Playing with these equations, once we obtained a couple of values for the shift and renormalization parameters, allows to obtain new values for the parameters appearing in these equations, which could then be used in \textsc{ISHEM} to simulate SEDs with the right scale and break frequency position. The reported equations represent an extension of equations (1) and (2) reported by \cite{Peault2019} to the non-conical geometry with also a more realistic angle dependence of the jet emission which reflects also  the improvements in the new version of {\sc ishem} used in the present work. A full derivation of these scaling relations is presented in Appendix~\ref{sec:scaling}. Finally, since the spectral emission from the source is largely dominated by the accretion flow beyond the optical wavelengths, we were not able to constrain the "cooling break" of the spectrum, which is expected at high energies \citep[see, e.g][]{Peer2014}. We then assume the jet optically thin synchrotron emission to extend with a power-law shape at least up to the hardest X-ray bands of the observed SED. We note that the extrapolation of the observed IR power-law spectrum at high energies implies that the jet has a negligible contribution in the hard X-ray band (see Subsection \ref{ss:global}).
\section{Spectral Analysis}
In NS Low Mass X-ray Binaries the jet emission is expected to dominate only the radio-to-IR wavelengths, while the emission from optical to X-ray should be mainly ascribed to the accretion disk (since usually the radiation emitted by the faint companion is negligible). Therefore, we began with a separate analysis for the radio-to-IR data, fitted with \textsc{ish}. Then the optical-to-X-ray data were described mainly with \textsc{diskir}, an irradiated disc plus Comptonization model \citep{Gierlinski2008}. However, although X-ray reprocessing from the outer disk or even direct emission from the outer disk is expected to dominate the Near Infrared (NIR) - optical region in NS LMXBs \citep{Russell2006,Russell2007}, some level of contribution from the jet emission might still be present \citep[several examples can be found in, e.g.][]{Lewis2010,Harrison2011,Baglio2016,Baglio2019}. We therefore performed a fit of the whole data set, in order to check if accretion and ejection do dominate over two separate frequency ranges or either if there is a border territory, i.e. the NIR-optical region, where these phenomena can not be easily singled out and have to be taken into account together.
\subsection{From radio to IR: the jet emission}\label{ss:jetem}
\begin{figure}
\centering
\includegraphics[scale=0.37]{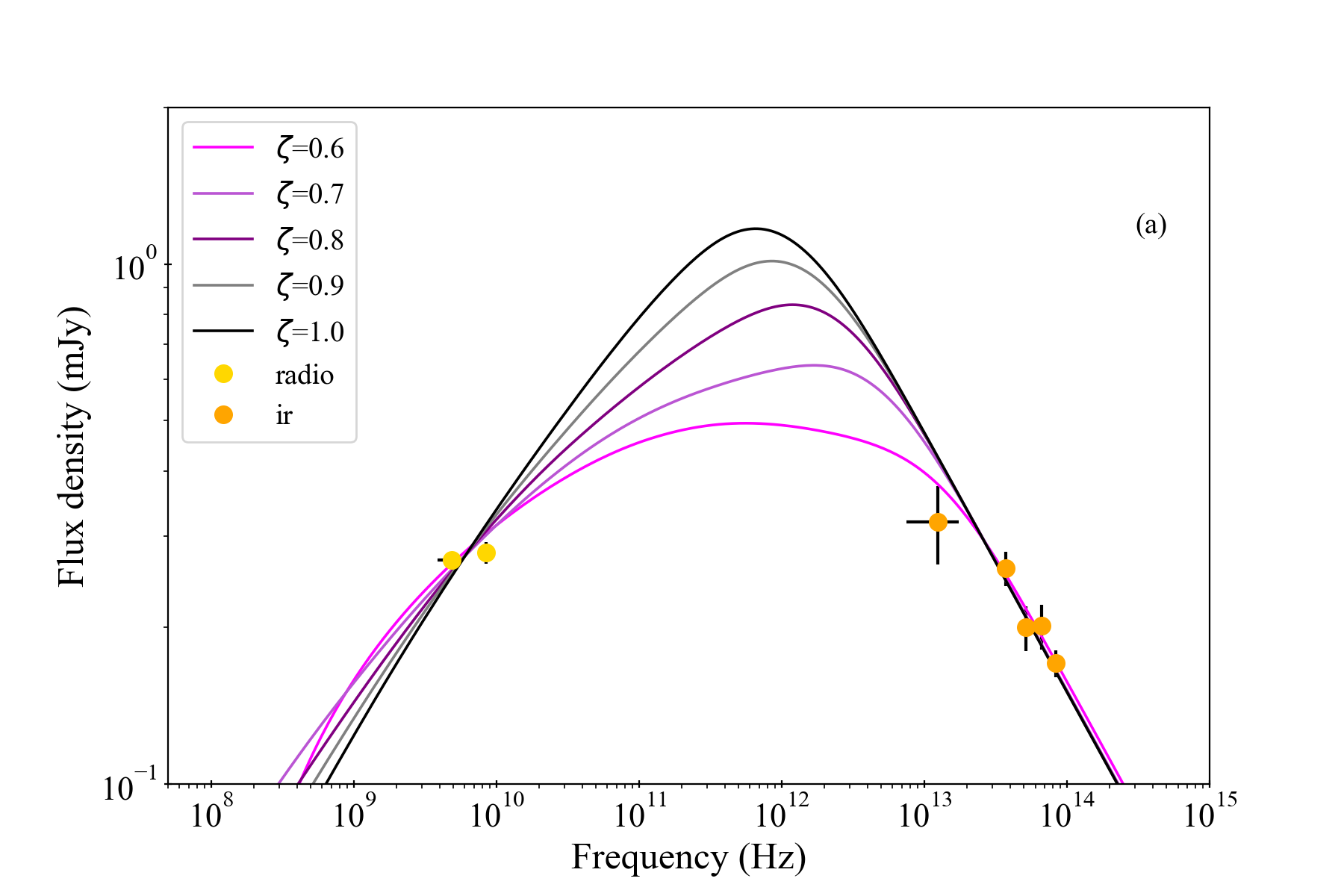}
\includegraphics[scale=0.37]{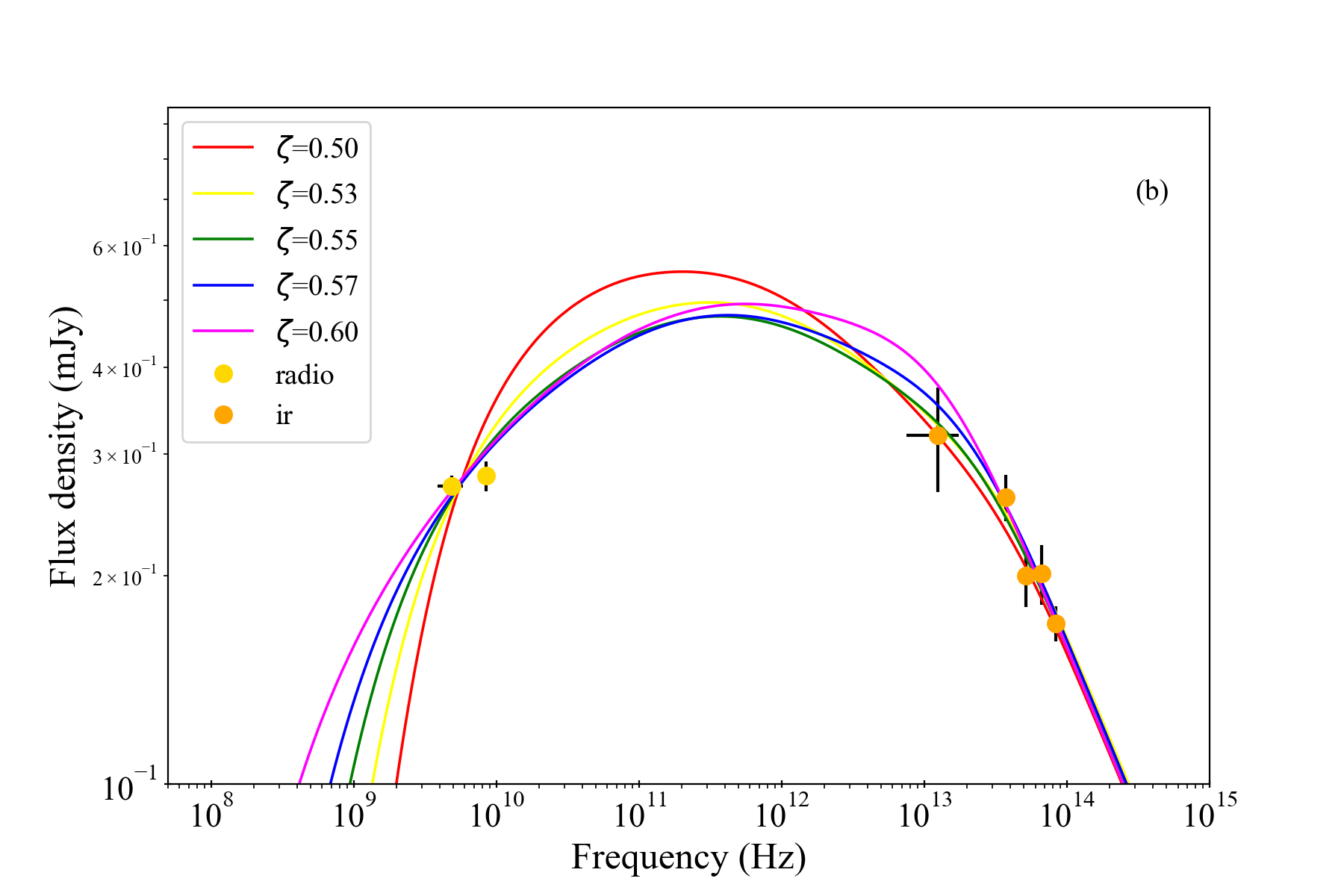}
\caption{Simulated SEDS with ISHEM using several values of $\zeta$, from 1.0 to 0.6 (panel \emph{a}) and in the critical region 0.5 to 0.6 (panel \emph{b}), normalized and shifted in order to fit the data. In all these models, the X-ray PDS was used as an input for ISHEM. Panel \emph{b} shows the interesting evolution of the SED for values of $\zeta$ spanning in the crucial region between $\zeta=0.6$ and $\zeta=0.5$: when the jet becomes too collimated, the contribution from the external regions of the jet (the lower frequencies) becomes dominant and it leads again to an inverted spectrum.}
\label{fig:geometries}
\end{figure}

We first tried to reproduce the observed SED with a standard set of parameters, which are listed in Table \ref{tab:ishem}, with $p=2.0$, the X-ray PDS and adopting the usual conical geometry ($\zeta$=1). We chose $p=2.0$ according to the fit to the optically thin part of the jet spectrum by \cite{Migliari2010} and also to the standard diffusive shocks acceleration theory. We then tested the "synthetic" spectrum on \textsc{Xspec}, using the \textsc{ish} model built on it to fit the data. We also checked if using a one-Lorentzian model instead of a double-Lorentzian model, i.e. ignoring the component for the high frequency QPO (see Subsection \ref{ss:timing}) which could probably be related to the orbital motion of the system \citep{Stella1998}, could influence the results of the fit. We found that both models lead to the same results, therefore in the following we will refer only to the results obtained including the QPO. \\
Even if IR fluxes are expected to be only slightly affected by interstellar reddening, we included the model \textsc{redden}, which estimates the extinction in the optical band, $E(B-V)$. The latter was frozen to 0.5 (see Sec. \ref{ss:diskemi}), since it was left unconstrained by the fit. The outcome of the fit is quite poor, as witnessed by the resulting $\chi^2_\nu$ (d.o.f.) of 3.96 (5). 
Furthermore, using Equations \ref{eq:fluxbr}-\ref{eq:break} to explore the parameters needed to improve the simulation, we found that, in order to have reasonable values for the jet power of the expected order of magnitude, i.e. 0.01 $L_{\rm Edd}$, one has to invoke oddly high opening angles (see Section \ref{sec:disc}). As mentioned in Section \ref{sec:ishem} the shape of the SED can be affected by only three elements: the shape of the electron distribution (which modifies the slope of the optically thin part of the spectrum), the geometry chosen and the dissipation pattern of the ejecta, in this case based on the X-ray PDS. Since the IR data are well fitted by the optically thin region of the synthetic SED, the chosen value of $p$ seems to be correct, as expected. In the following we will then try to change the geometry first and the PDS then in order to see if, with a different choice for these ingredients, we can still find a good model for the data. 
In order to check if a different value for $\zeta$ might improve the results of the fit, we run again the simulations with different $\zeta$ between 0.5 and 1.0, and we repeated the whole procedure. 
The resulting best fits as a function of $\zeta$ is shown in Figure \ref{fig:geometries}, (\emph{a})-(\emph{b}). 
The results of each fit is reported in Table \ref{tab:geometries}. The values of $\zeta$ for which we have the lowest $\chi^2_{\nu}$ are 0.57 and 0.6 (1.08 and 1.13 respectively, both with 5 d.o.f.). A paraboloidal jet with $\zeta$ in this range of values therefore represents an acceptable fitting scenario, contrary to the typical conical geometry. \\  As mentioned in Subsection \ref{ss:param}, for strongly non-conical geometries, spectra obtained with longer simulation times 
could produce flatter spectra. We show in Table \ref{tab:tsimu} the results of the fits for three values of $\zeta$, i.e. 0.53, 0.60 and 0.7, with $t_{\rm simu}=1\times 10^5$ and $t_{\rm simu}=3\times 10^5$. As expected, the values of $\chi^2_\nu$ are generally lowered by increasing the simulation time, with the exception of the fit with $\zeta=0.7$, which is mostly unaffected by changing $t_{\rm simu}$. Therefore we found out that even with higher simulation times, the best fit is obtained again for $\zeta$ around 0.6. While an even higher simulation time would still be physically acceptable (see Subsection \ref{ss:shape}), it would very likely only confirm the results presented here with shorter and more feasible $t_{\rm simu}$. \\ 
We also notice that our conclusion on the best value of $\zeta$ should not be considered conclusive, as further investigations in the range between $\zeta=$0.53 and $\zeta=$0.6, with possibly higher $t_{\rm simu}$ could, in principle, lead to even more accurate estimates of the best $\zeta$. However, a precise estimate of $\zeta$ goes beyond the scopes of this work and would likely provide no or very little improvement to the results.

We therefore conclude that a non-conical geometry, with $\zeta$ around 0.6 and possibly even below, improves significantly the fit with \textsc{ISHEM} using the X-ray PDS. \\

\begin{figure*}
\centering
\includegraphics[scale=0.60]{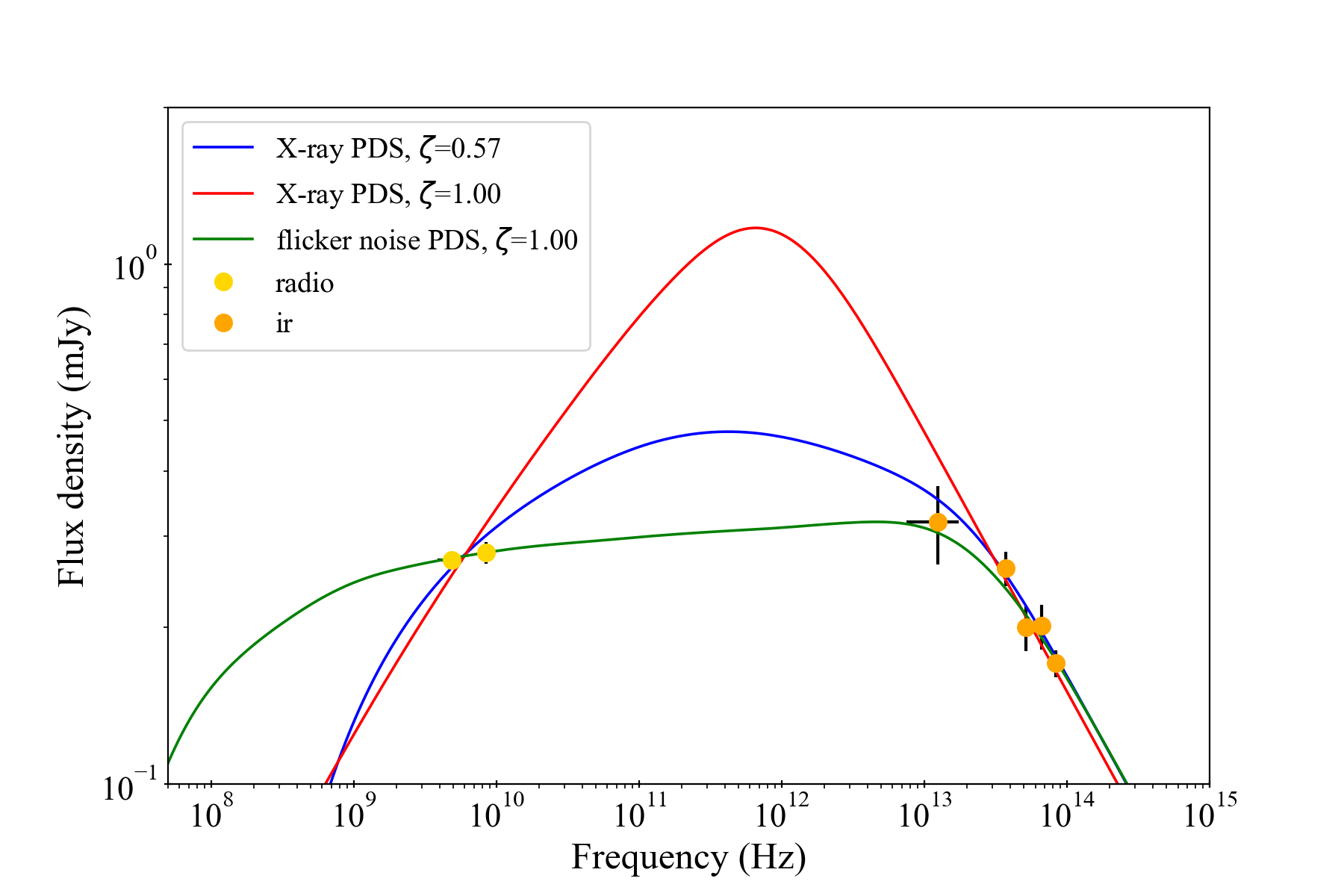}
\caption{Simulated SEDS with ISHEM for varying geometries and input PDS as compared with the radio-to-IR data set.}
\label{fig:final}
\end{figure*}

\begin{table*} 
\centering
\caption{Results of the \emph{VLA}-\emph{Spitzer}/IRAC fits with \textsc{ISH} with different $\zeta$ values. In all the fits, the number of degrees of freedom is equal to 5.}
\begin{tabular}{l c c c c c c c c c c}
\hline 
\hline
\multicolumn{10}{c}{\textbf{Fit results for different jet geometries}} \\
\hline
& \multicolumn{9}{c}{$\zeta$} \\
\cmidrule(lr){2-10}
& 0.50 & 0.53 & 0.55 & 0.57 & 0.60 & 0.70 & 0.80 & 0.90 & 1.00 \\
 {${\bf \chi^2_{\nu}}$} & 5.95 & 2.34 & 1.46 & 1.08 & 1.13 & 2.11 & 2.84 & 3.37 & 3.96 \\  
\hline
\hline
\end{tabular}
\label{tab:geometries}
\end{table*}

\begin{table} 
\centering
\caption{Results of the \emph{VLA}-\emph{Spitzer}/IRAC fits with \textsc{ISH} with different $\zeta$ and $t_{\rm simu}$. In all the fits, the number of degrees of freedom is equal to 5.}
\begin{tabular}{l  c c | c c | c c  }
\hline 
 \multicolumn{7}{c}{\textbf{Fit results for different jet geometries and simulation times}} \\
\hline
& \multicolumn{6}{c}{$\zeta$} \\
\cmidrule(lr){2-7}
& \multicolumn{2}{c}{0.50} & \multicolumn{2}{c}{0.60} & \multicolumn{2}{c}{0.70} \\
\cmidrule(lr){2-7}
$t_{\rm simu}$  ($\times 10^5$ s) &  1 & 3 & 1 & 3 & 1 & 3 \\ 
\cmidrule(lr){2-7}
 {${\bf \chi^2_{\nu}}$} &  2.34 & 1.36 & 1.13 & 0.80 & 2.11 & 2.13 \\  
\hline
\hline
\end{tabular}
\label{tab:tsimu}
\end{table}
We then tested the other possible scenario, where the observed X-ray variability does not reflect the fluctuations of ejection  velocity, using a "flicker-noise" PDS in conical geometry. We assumed a rms fractional amplitude of 30\% and included a range of frequencies ranging from $f_1=10^{-5}$ Hz to $f_2=10^{3}$ Hz. Using the same set of parameters shown in Table \ref{tab:param}, we obtained a synthetic SED which is in quite good accordance with the data, i.e. $\chi^2_\nu$ (d.o.f.)=0.27(5). The best-fit model is shown in Figure \ref{fig:final}, in comparison with the data and a pair of best-fit models obtained with the X-ray PDS and variable values of $\zeta$.
\subsection{From optical to X-ray: the disk emission}
\label{ss:diskemi}
While in the previous section we focused on the part of the SED dominated by the jet, in this section we will focus on the optical-to-X-rays data, which are expected to be dominated by the disc and the hot corona emission. \\ 
First of all we used \textsc{diskir} \citep{Gierlinski2008}, which includes the disk emission, Comptonization from a hot corona and the X-ray illumination of the disk, relevant in  the presence of data coverage in the optical-UV domain (as in our case). The main parameters of the model are: the temperature of the disk at its inner radius $kT_{\rm disk}$, the $\Gamma$ index of the power-law reproducing the Comptonization spectrum, the electron temperature of the corona $kT_{\rm e}$, the ratio $L_{\rm C}/L_{\rm D}$ between the luminosity of the Comptonized emission and the disk luminosity, the fraction of the flux of the Compton tail which is thermalized in the inner and outer radius ($f_{\rm in}$ and $f_{\rm out}$ respectively), the radius of the illuminated disk $r_{\rm irr}$, the outer disk radius $\log{r_{\rm out}}$ (both in units of the inner disk radius) and the normalization $K$, which can be used to derive the inner disk radius.\\
The value $L_{\rm C}/L_{\rm D}$ is generally used as an indicator of the spectral state and it is usually higher than 1 in intermediate/hard states. In the following fits we fixed $f_{\rm in}$ and $r_{\rm irr}$ to the standard values of, respectively, 0.1 and 1.1 \citep{Gierlinski2008}. Furthermore, as 4U 0614+091 is known to be an Ultra-Compact Binary, we fixed $\log{r_{\rm out}}$ to the value of 3. This value is reasonable considering that, since the system has an orbital period of about 50 mins, the orbital separation is expected to be around 3$\times 10^{5}$ km and a factor of 10$^3$ guarantees that even with large inner disk radii, say 100 R$_{\rm G}$, as expected in hard state, it is larger than the extension of the disk. We also included a black-body model in the spectrum (\textsc{bbody} in \textsc{Xspec}), already present in the fit by \cite{Migliari2010}, which accounts for the emission of the NS surface (or the boundary layer). Finally, we included the two Gaussian components used by these authors, i.e. at 0.67 keV for the O VIII line and at 6.6 keV for the Fe K fluorescence line. We applied to the model the components \textsc{redden} and \textsc{tbabs} to take into account interstellar extinction in both the UV and X-ray band. Finally, we included \textsc{constant} to serve as a cross-calibration constant and we checked that its value was always around 1, i.e. in the range 0.8-1.2. \\
The fit did not constrain $L_{\rm C}/L_{\rm D}$, which was therefore tentatively fixed to 10, which is a reasonable value for hard states \citep[see, e.g.][]{DelSanto2008}. The fit is unable to constrain the parameters of the \textsc{gaussian} component at 6.6 keV, due to a marginal contribution of this component to the fit, i.e. a 9\% probability of improvement by chance (calculated via \textsc{ftest}). This is not surprising, since this is not the first time that the iron line in 4U 0614+091 was found very weak \citep[see, e.g][]{Piraino1999}. In the following we will therefore not include this \textsc{gaussian} component. On the other hand, we confirmed the presence of a broad line at $\sim$ 0.67 keV, likely associated to O VIII (see Section \ref{ss:source} for references). The fit provides E(B-V) (from \textsc{redden}) of about 0.5, which corresponds to a value\footnote{Keeping in mind the relation $A_{\rm V} =R_{\rm V}\times E(B-V)$, with $R_{\rm V}$ fixed to 3.1 \citep{Seaton1979b,Seaton1979a};} of A$_{\rm V}$ slightly lower than the one reported in \cite{Migliari2010}, equal to 2. We were able to find only a relatively high lower limit to the corona temperature, i.e. 110 keV, which might be probably due to a the lack of a proper modeling of the high energy hard tail \citep[see, e.g][ and references therein]{DiSalvo2001,Iaria2001, Dai2007,DelSanto2013} rather than such a high temperature electron plasma. The results of the fit led as well to a significantly colder disk with respect to \cite{Migliari2010}, i.e. $kT_{\rm disk} < 0.1$ keV, but correlations with the disk normalization $K$ and/or with the imposed values of $L_{\rm C}/L_{\rm D}$ might be at play. The normalization of the disk $K$ is bound to the inner radius of the disk $R_{\rm in}$ by the relation: $K=(R^2_{\rm in}/D^2_{\rm 10 kpc})\times \cos{i}$, with $D_{\rm 10 kpc}$ is the distance of the system in units of 10 kpc. We found an apparent inner disk radius of $\sim$ 160 km ($\sim$ 73 R$_{\rm G}$), which has to be taken as a lower limit since it does not take into account the proper (unknown) inclination of the system and the correction factor \citep[for a more detailed calculation of the inner disk radius based on $K$ see, e.g.][ and references therein]{Marino2019b}. The few differences in our results with respect to the results obtained by \cite{Migliari2010} on the same X-ray data are likely due to either the inclusion, in our data set, of the 
SMARTS-UVOT data or to the different models used in the two papers.
\begin{table}
\centering
\caption{Fit results of the disk-dominated SED region, with data from SMARTS, UVOT, XRT, PCA and HEXTE. Quoted errors reflect 90\% confidence level. The parameters which were kept frozen during the fits are reported between round parentheses.}
\begin{tabular}{l  l  l l}
\hline 
\hline
\multicolumn{4}{l}{\textbf{Spectral analysis}} \\
\hline
 \textsc{redden} & \textbf{E(B-V)} & & 0.48$^{+0.07}_{-0.06}$\\
 \textsc{tbabs} & \textbf{N$_H$} & $\times$10$^{22}$ cm$^{-2}$ & 0.21$\pm$0.02\\
\multirow{9}{*}{\textsc{diskir}}& \bf{$kT_{\rm disk}$} & keV & 0.077$\pm$0.003\\
& \bf{$kT_{\rm e}$} & keV & >110\\
& \bf{$\Gamma$} & & 2.24$\pm$0.02\\
& \bf{$L_{\rm C}/L_{\rm D}$} & & (10)\\
& \bf{$f_{\rm in}$} &  & (0.10) \\
& \bf{$f_{\rm out}$} & ($\times$10$^{-3}$)  & 2.0$^{+9.0}_{-1.3}$ \\
& \bf{$R_{\rm in}/\sqrt{\cos{i}}$} & ($R_{\rm G}$) & 73$^{+5}_{-4}$ \\ 
& \bf{$R_{\rm irr}/R_{\rm in}$} & & (1.01)\\
& \bf{$R_{\rm out}/R_{\rm in}$} & & (10$^3$) \\ 
\multirow{2}{*}{\textsc{gaussian}} & \bf{$E_{\rm line}$} & (keV) & 0.671$\pm$0.009\\
& \bf{$\sigma$} & (keV) & 0.077$\pm$0.006\\
\textsc{bbody} & $\bf{kT_{\rm bb}}$ & (keV) & 1.38$\pm$0.02\\
\hline
& & \multicolumn{2}{l}{$\bf{\chi^2_{\nu}}(d.o.f.)$ = 1.49(442)}\\
\hline
\hline
\end{tabular}
\label{tab:fit}
\end{table}
\subsection{Global multi-wavelength analysis} \label{ss:global}
\begin{figure*}
\centering
\includegraphics[scale=0.34]{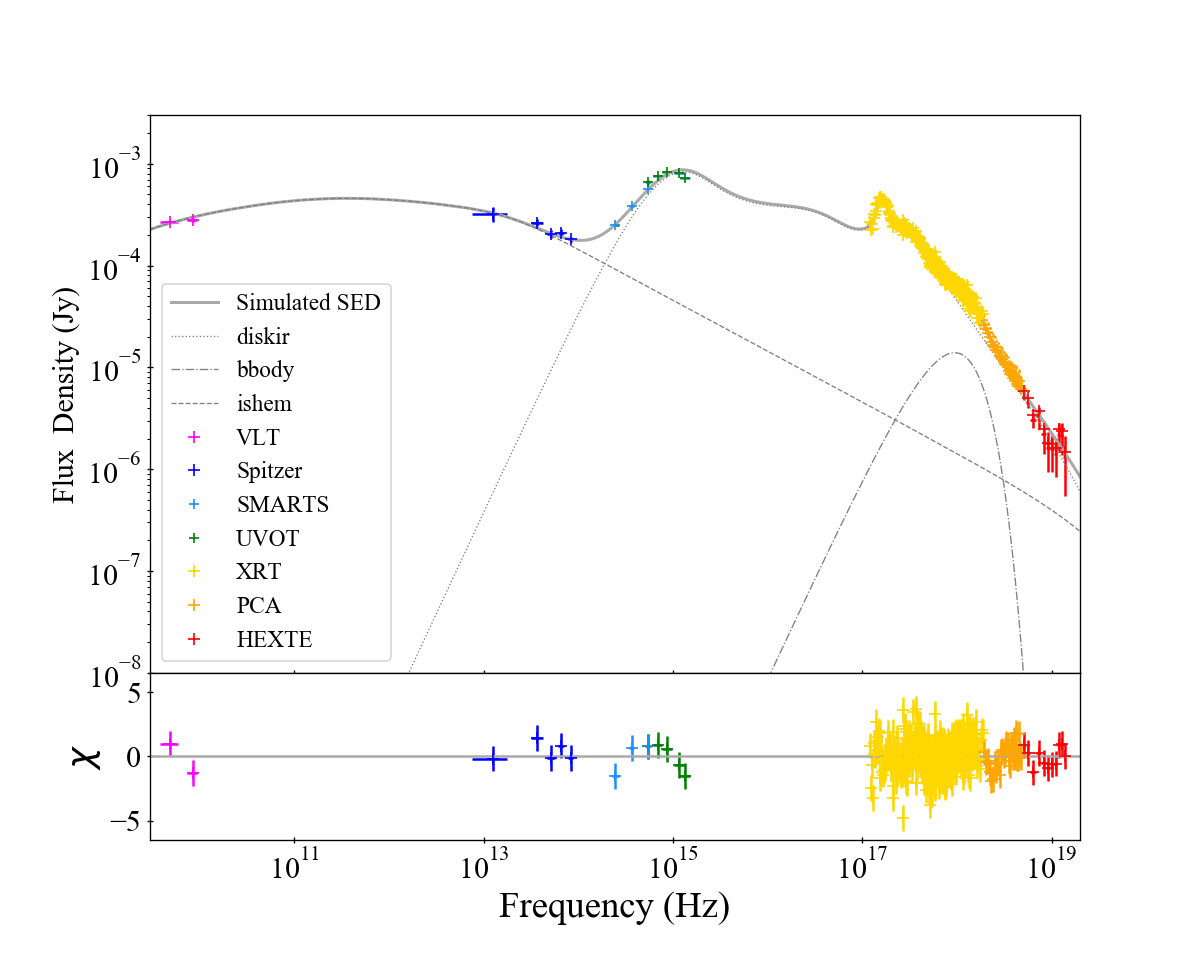}
\includegraphics[scale=0.34]{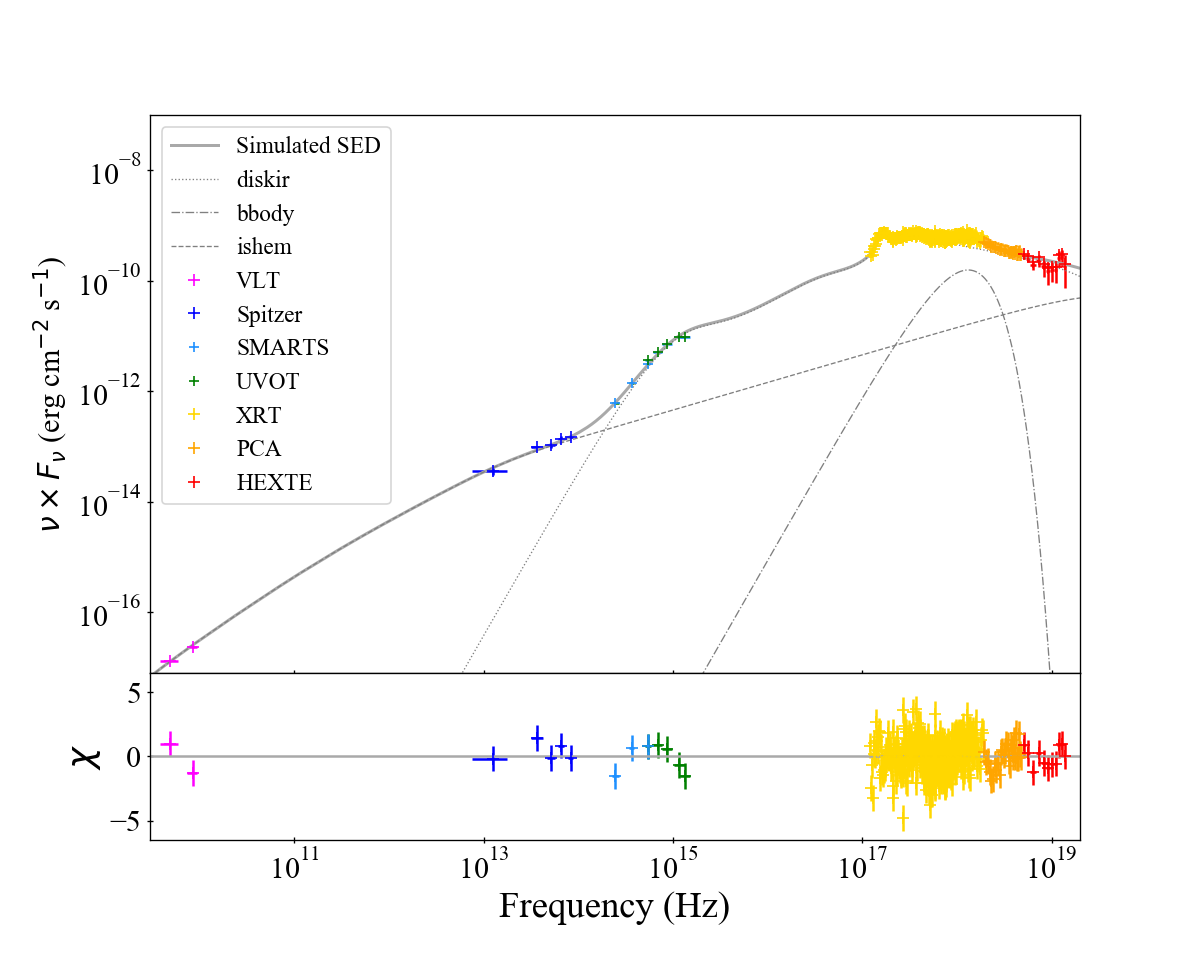}
\includegraphics[scale=0.34]{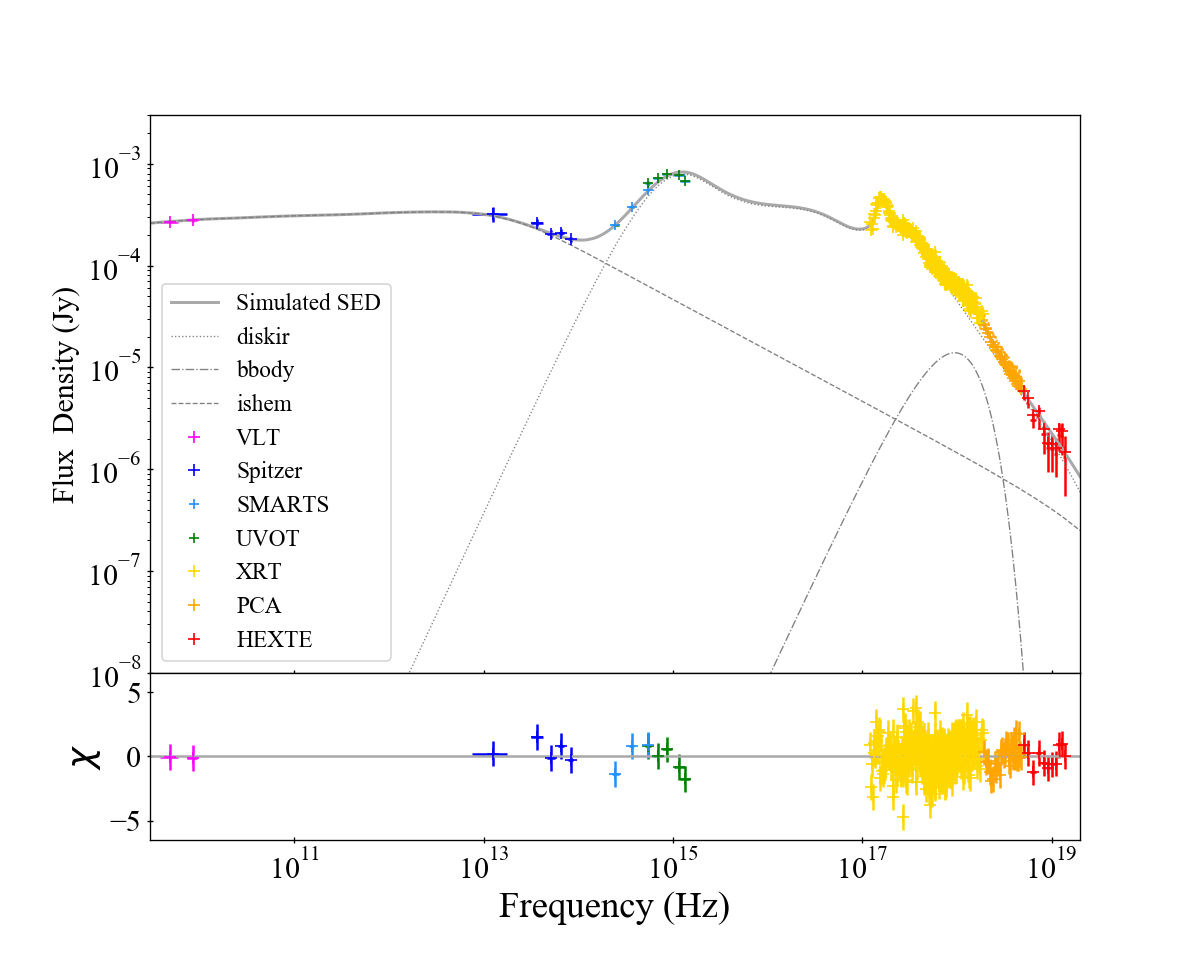}
\includegraphics[scale=0.34]{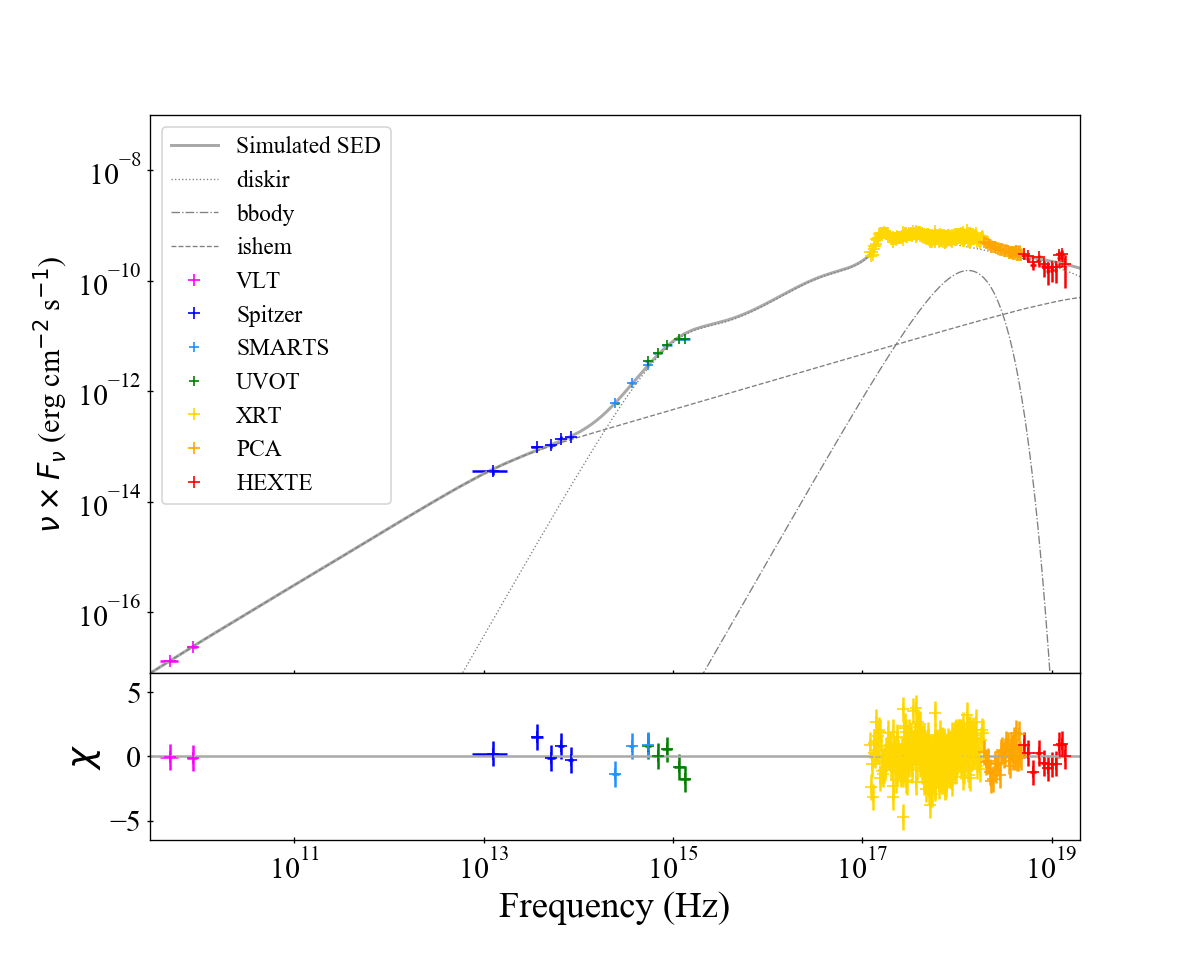}
\caption{Best-fit unabsorbed spectral energy distribution compared with the whole multiwavelength dataset available for 4U 0614+091, in both Flux Density (\emph{Left}) and $\nu\times F_\nu$ (\emph{Right}) representation. The optical-to-X-ray data set has been analysed with an irradiated disk + black-body model, while the radio-to-IR data set was modeled with ISHEM. In this plot we show both best-fit ISHEM models found in this paper: on \emph{Top} the model built with the X-ray PDS and corresponding to a non-conical geometry ($\zeta=0.6$) and on (\emph{bottom}) the model corresponding to a "flicker noise" PDS in conical geometry.}
\label{fig:fit}
\end{figure*}

The spectral analyses conducted in subsections \ref{ss:jetem}-\ref{ss:diskemi} allowed us to characterize separately the emission from the jet and from the disk, under the hypothesis that their spectral domains were independent. In the following we report on the global dataset fitted with both accretion and ejection models, in order to confirm the previous assumption and provide a final, multi-wavelength study of the spectral energy distribution of 4U 0614+091. \\
Starting from the best-fit model for the optical-to-X-ray data, whose main parameters are reported in Table \ref{tab:fit}, we included the \emph{VLA} and \emph{Spitzer} data used in Subsection \ref{ss:jetem} and added \textsc{ish} to the spectral model employed. Since it is not possible to exclude \emph{a priori} that our results on the jet might have been biased by the lack of higher frequencies data, we performed several fits trying different ISHEM models with $\zeta$ spanning from from 1.0 to 0.5. Unfortunately each fit results in approximately the same $\chi^2_{\nu}$ (d.o.f.) value of 1.49(451). Similarly, using the ISHEM model based on the "flicker noise" PDS, the fit results in a $\chi^2_{\nu}$ (d.o.f.) value of 1.43(451).  
This situation is not surprising because the fit is indeed dominated by the higher statistics data in the X-ray and only slightly affected by the modeling of the few data points in the radio-IR domain. Leaving the same model, i.e. including \textsc{diskir} and \textsc{bbody}, but neglecting all the data but radio and IR, results in fits which are similar to the fits performed in Subsection \ref{ss:jetem}: among the different fits with X-ray PDS with varying $\zeta$, the best fit is obtained again with $\zeta=0.57$, i.e. for which the fit goes from $\chi^2/$d.o.f.=671.9/451 (including optical-to-X-rays data) to $\chi^2/$d.o.f.=5.25/5., while choosing the flicker noise PDS the fit goes from $\chi^2$/d.o.f.=644.9/451 to a $\chi^2$/d.o.f. of 3.55/5. However, the extension of the data set does have some effects on the ISHEM best-fit parameters, i.e. the shift frequency and re-normalization factors, which are different with respect to the previous set of fits. In the next section we use Equations \ref{eq:fluxbr}-\ref{eq:break} in order to check whether the best fit shifts and normalization allows for 'reasonable' physical parameters of the jet.
Finally, we checked if the results are dependent on the values assumed by the cross-calibration constant for the IR and the radio data, which were assumed to be equal to 1.0 for the radio and IR data; leaving the cross-calibration parameters free to vary between 0.8 and 1.2 does not change significantly the values obtained for the $\chi^2_{\nu}$ (d.o.f.). In the following we will therefore only refer to the fits with the parameters fixed to 1.0. \\
We refer to Figure \ref{fig:fit} for the SED, overimposed to the best-fitting ISHEM model. The best-fit parameters of the accretion flow model found with the lower frequencies extension of the dataset are all perfectly compatible with the results reported in Table \ref{tab:fit}. 

\section{Discussion} \label{sec:disc}

\begin{figure*}
\centering
\includegraphics[scale=0.45]{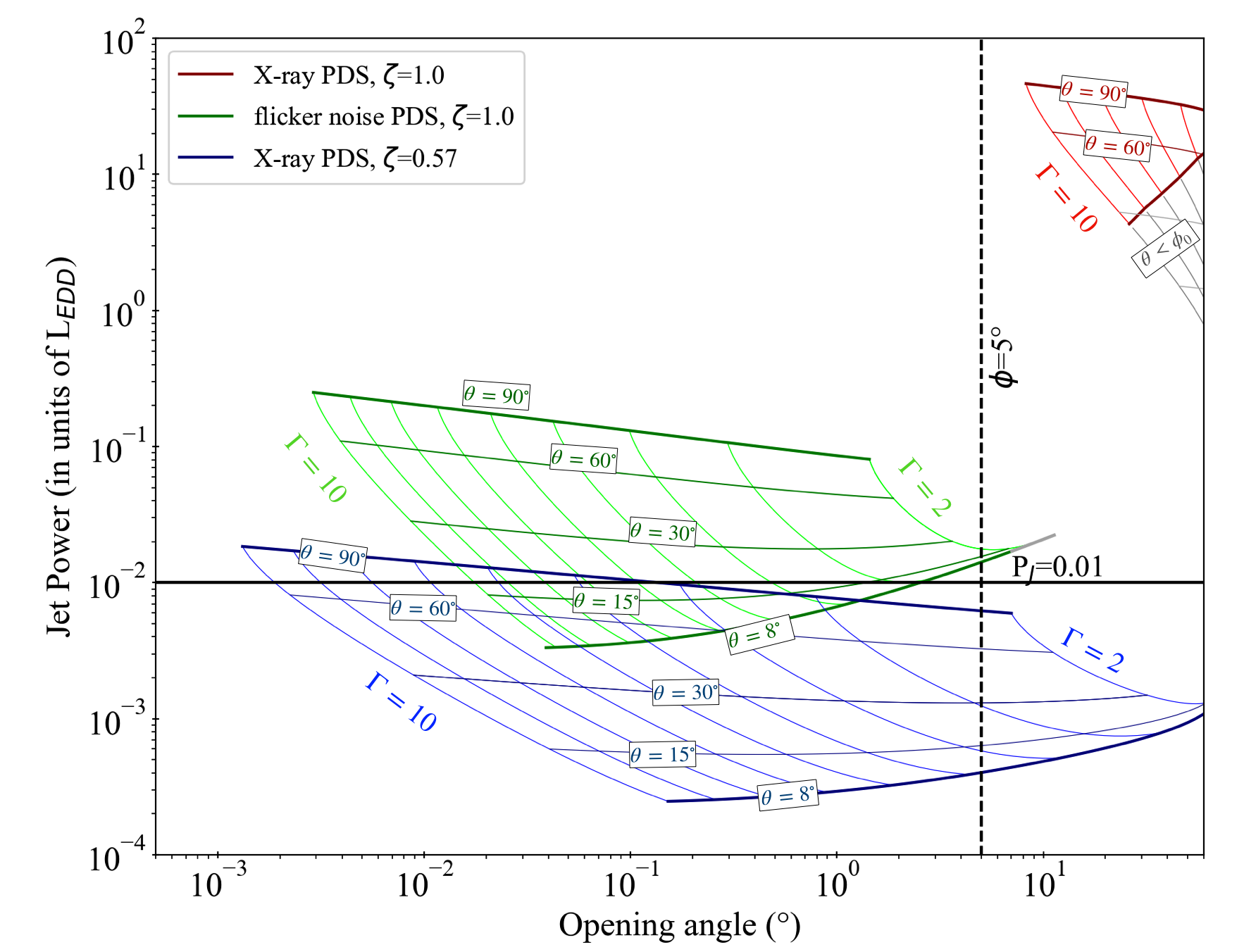}
\caption{Jet power - (observed) opening angles curve (from Equations \ref{eq:fluxbr}-\ref{eq:break}) for a different combinations of $\zeta$ and PDS to a range of possible values for the inclination $\theta$ and a range of $\Gamma_{\rm av}$ between 2 and 10. In particular, in each "region" (identified by a specific color) the inclination increases going upwards, while the Lorentz factor increases going from right to left. Curves for specific fixed values of $\Gamma$ and $\theta$ are identified to help the eye. In each area, the sub-region for which the condition $\theta>\phi_0$ is not satisfied are colored in grey and have to be discarded, as the used scaling relations are no longer valid}. In this plot, the black solid line indicates the expected jet power, while the dashed black line points out the upper limit for the opening angle.
\label{fig:areas}
\end{figure*}

\begin{figure}
\centering
\includegraphics[scale=0.30]{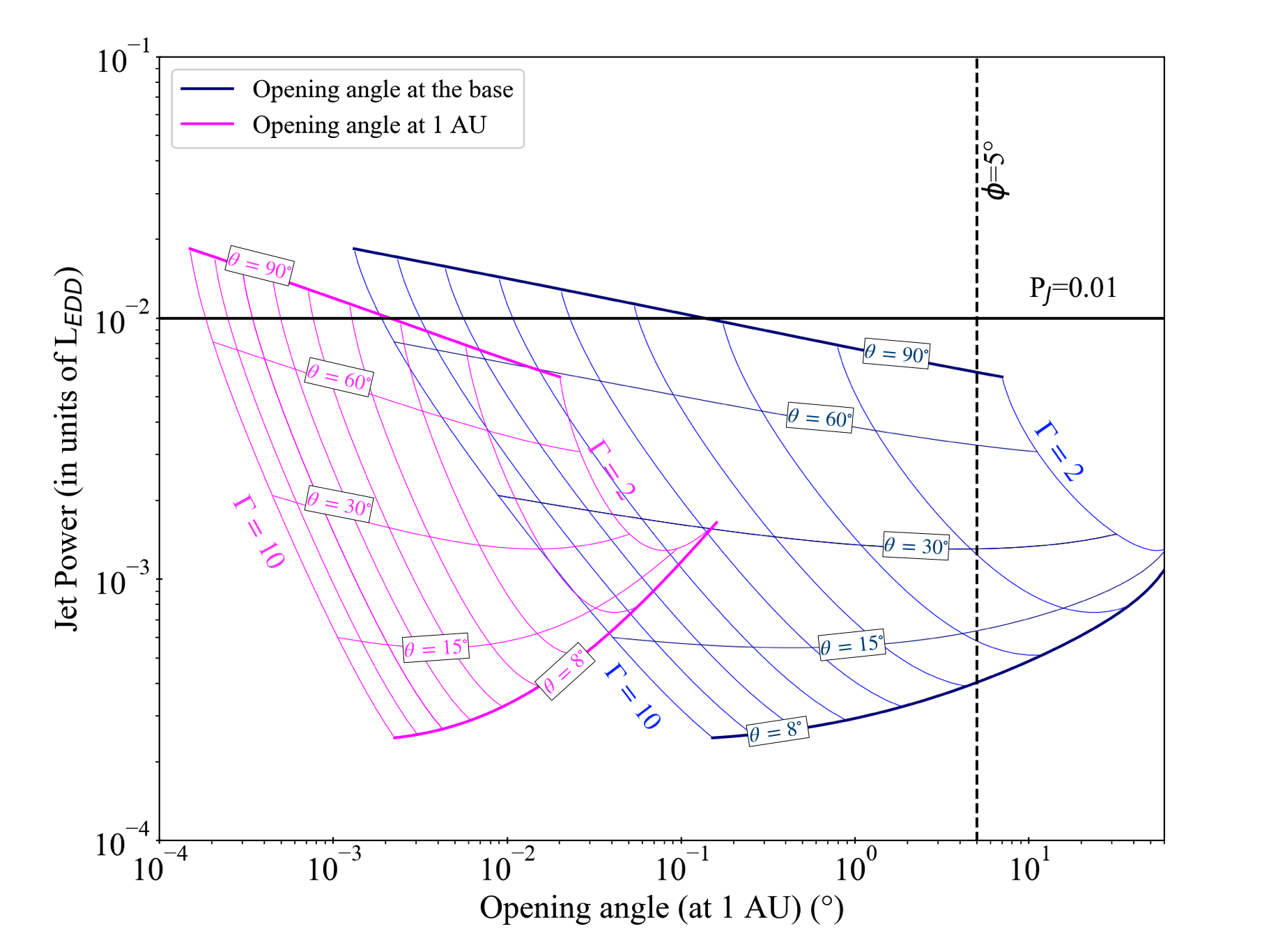}
\caption{Jet power - opening angle curves for the particular case of $\zeta=0.57$, using the opening angles at the base of the jet (blue) and the opening angles hypothetically observed at 1 AU (magenta). We refer to the caption of Fig. \ref{fig:areas} for further details.}
\label{fig:angles}
\end{figure}

The analysis limited to the radio-IR domain carried out in Subsection \ref{ss:jetem} suggests two possible scenarios to describe the jet emission for 4U 0614+091 within the internal shocks scenario: on one hand the variability in the Lorentz factors of the ejecta is related to the X-ray variability, i.e. the variability in the accretion flow, but the jet is non-conical (in the following scenario \emph{a}), on the other hand it is also possible that, at least in this source, a flicker noise power spectrum, unrelated to the X-ray variability, is a better proxy for the fluctuations of the jet Lorentz factor (scenario \emph{b}). Including the optical-to-X-rays portion of the data set in Subsection \ref{ss:global} turned out to be furthermore inconclusive in discerning between the two proposed scenarios. This is mostly due to the lack of data around 10$^{11}$-10$^{12}$ erg s$^{-1}$. The availability of data in this region, e.g. from the Atacama Large Millimeter/submillimeter Array (ALMA), would have been crucial to distinguish between the two scenarios. For example, ALMA data for other NS LMXBs in the hard state \citep[see, e.g., the SEDs shown by  ][]{DiazTrigo2017,DiazTrigo2018} indicate a small rise in flux density in this region which, if observed also in our data set, would have likely favored scenario \emph{a}. \\
We will instead explore and discuss both scenarios \emph{a} and \emph{b} in the following section, on the basis of the global fit reported in Subsection \ref{ss:global}. In both cases there is evidence that it is not possible to completely disentangle the disc contribution to the IR domain or the jet contribution to the optical domain as well. In particular, it results that there is a jet contribution varying from 30\% to 6\% in the \emph{SMARTS} wavelengths range within scenario \emph{a}, while this contribution is less prominent in scenario \emph{b}, i.e. from 20\% to 3\%. These results confirm the study led by \cite{Russell2006}, according to which in NS LMXBs the main emission process to be taken into account in the optical domain is the X-ray reprocessing from the disc\footnote{Contrarily to black holes, where the jet emission is usually extended until the optical wavelengths \citep[see, e.g][]{Peault2019}}, even though they also point out how the jet contribution might not be negligible at all.  
\subsection{Scenario \emph{a}: a non-conical jet?}
 Due to the high statistics in the X-rays, the fit to the whole dataset gives comparable $\chi^2_\nu$ values for each of the tested geometries. In order to determine the most likely values of $\zeta$ we use Equation \ref{eq:fluxbr}-\ref{eq:break} to convert the values of the shift and renorm factors found by \textsc{ish} in couples of $P_{\rm J}$-$\phi$. We explored also how these results were influenced by our choice for $\theta$ and $\Gamma_{\rm av}$, allowing for both to change in some physically reasonable ranges. In particular, since the scaling relations used in this work, i.e. Eq. \ref{eq:fluxbr}-\ref{eq:break}, are valid in the approximation of $\phi_0 \lesssim \theta$ (see \ref{localcylapp}), with $\phi_0$ the opening angle at the base of the emitting region, we fixed the lower limit for the inclination $\theta$ to 8$^\circ$. On the other hand, $\Gamma_{\rm av}$ was allowed to vary in the 2-10 range. Running these tests draw areas of possible results in a $P_{\rm J}$-$\phi$ plot, corresponding to specific values of $\zeta$.
 We show in Figure \ref{fig:areas} the resulting area for $\zeta=1.0$ (red) and $\zeta=0.57$ (blue).
 In each of the resulting skewed areas, the bottom of the areas corresponds to $\theta=8^\circ$, the top corresponds to $\theta=90^\circ$, while $\Gamma_{\rm av}$ increases from right (where $\Gamma_{\rm av}=2$) to left ($\Gamma_{\rm av}=10$). The sub-regions where the condition $\theta<\phi_0$ are colored in grey and they have to be excluded. It is important to notice that for a non-conical geometry, the opening angle depends on distance $z$ along the jet. The emitting region, located at about 1 AU from the base of the jet, will have therefore a different opening angle than the values encompassed by the blue area in Fig. \ref{fig:areas}. Thereby this angle \emph{can} violate the condition $\theta<\phi$. Indeed, for a fixed geometry parameter $\zeta=0.57$, using the opening angle at a distance of 1 AU $\phi_0$ results in significantly smaller angles, i.e. below 0.1$^\circ$, as shown in Fig. \ref{fig:angles}. Such extreme values are not implausible, as small opening angles have been suggested for jets in XRBs \citep[see, e.g.][]{Zdziarski2016}. Both the ranges of jet powers and opening angles individuated by the $\zeta=0.57$ areas can be accepted and a $\zeta\approx$0.6 value appears still consistent with the best physically motivated scenario. \\
 For $\zeta=1.0$, on the contrary, the area found by this procedure does not allow for reasonably small opening angles and requires very high jet powers for the whole range of explored $\Gamma_{\rm av}$. This represents another point in favor of ruling out the conical geometry scenario. \\
The concept of a non-conical jet is not groundbreaking: it is known that the conical geometry is an approximation of the real geometry, as valid and efficient as it proved to be. The confinement agent necessary can be internal or external. In the first case, it has been proposed by several authors that collimation might be due to a toroidal component of the magnetic field which increases along the axis and that forces the jet to decrease its opening angle \citep[see, e.g. ][]{Heyvaerts1989,Pudritz2006,Pudritz2012}. However, this mechanism has been questioned by \cite{Spruit2010}, according to which a magnetic self-confinement of the jet is not physically possible as the toroidal magnetic pressure within the jet would force them to expand. On the other hand the collimation necessary to "break" the \citep[unstable,][]{Marti2016} conical geometry might be exerted for instance by the interstellar medium \citep[e.g.][]{Asada2012} or an external magnetic field kept in place by the disk \citep[e.g.][]{Spruit1997}.
 
It is interesting to compare these results with those previously found by applying the same ISHEM model to jets in BH-XRBs, where the conical geometry assumption worked correctly. In a few cases \citep[see for example the radio residuals for some spectra in ][, figure 3]{Peault2019} it is reasonable that flatter models could even improve the already acceptable fits and in this sense a non-conical geometry could be necessary. Any comparison between NS and BH XRBs is therefore premature for at least two reasons: a non-conical geometry was not tested for BH XRBs and, also, we need to test more NS LMXBs to draw any conclusion on a possible difference between jets in these two classes of systems.
\subsection{Scenario \emph{b}: X-ray variability is not a good proxy for the Lorentz factor fluctuations}
In the second scenario, the dissipation pattern of the shells internal energy in the jet is not related to the X-ray timing properties, i.e. the timing properties of the accretion flow, but it is mainly due to "flicker noise". The plausibility of this scenario is confirmed by the $P_{\rm  J}$-$\phi$ diagram in Fig. \ref{fig:areas}, since the corresponding area encompasses the expected range of jet powers-opening angle. \\ This is not groundbreaking either \citep[we refer again to, e.g.,][]{Jamil2010, Malzac2013} but it would be certainly different to the results obtained on the other sources to which the ISHEM model has been applied in the past. In this case the fact of having a NS instead of a BH might play a role. Under the hypothesis of a disk-jet coupling, the variability in the emission from the NS/boundary layer may not be transmitted to the ejecta in the jet, breaking subsequently the connection between the ejection pattern of the shells and the X-ray PDS. Alternatively, one might also consider differences in the jet launching mechanism in NSs with respect to BHs. For instance \cite{Parfrey2016} shows how the interaction between a fastly rotating low magnetized NS and the disk may lead to a state where the magnetic field lines are open and provide the energy necessary for the ejection of particles. In this case we do not expect that the dissipation pattern in the ejecta and the accretion flow fluctuations in the disk to be exactly matched. Such a mechanism could be at work in Accreting Millisecond X-ray Pulsars and analogous systems, which might possibly include 4U 0614+091. Indeed, the system, with a 415 Hz frequency spin \citep[see, e.g][ and references therein]{VanDoesburgh2017}, belongs to the family of binaries hosting millisecond NSs. However, the NS magnetic field is likely buried, as witnessed by the lack of observed X-ray pulsations, and this would make the attribution of the aforementioned mechanism to the system unlikely. \\
We also suggest the possibility that the lack of correlation between X-ray variability and ejecta in the jet proposed here for 4U 0614+091 may not necessarily be representative for the whole class of NS LMXBs in hard state. As apparent from the \emph{VLA} data used in this paper, the jet spectrum looks quite flat, which, we recall from Subsection \ref{ss:shape}, requires also an almost flat PDS. The X-ray PDS used has evidently not the required shape and it is therefore not surprising that it may not be the best tracer for the variability in the shells velocity. Similar PDS have been observed frequently in the so-called atoll LMXBs, when in island state (IS) \citep[see, e.g][]{VanStraaten2002, VanStraaten2003}. On the other hand PDS dominated by broad, flat-topped noise, similar to what observed in BH XRBs in hard state, have been found also in several atoll sources at low luminosity \citep{Belloni2002, Reig2004, VanStraaten2005}. This variability behaviour has been classified as Extreme Island State (EIS) \citep{Mendez1997,VanStraaten2003}\footnote{We refer the reader to figure 2 in \cite{Wijnands2017} for a direct comparison between the two types of PDS.} and systems in this state tend to populate a horizontal extension of the island state region in a color-color diagram, corresponding to harder spectra than sources in IS \citep{Muno2002, Gierlinski2002}. Indeed spectra of NS LMXBs in EIS are typically described with power-laws of $\Gamma \sim 1.8$ \citep[see e.g][]{Barret2000,Linares2008}, unlike e.g 4U 0614+091 which displays spectra usually steeper, $\Gamma \sim 2.2-2.4$, as in this paper and, e.g. \cite{Piraino1999,Migliari2010,Ludlam2019}. In addition, the X-ray variability is significantly stronger \citep[i.e. 30-40\% rms amplitude,][]{Wijnands2017} in EIS than sources in IS. The ensemble of these clues suggests that atoll sources in EIS are likely associated to a physical scenario where the disk is truncated far from the compact object and the NS surface is not very hot \citep{Reig2004,Bult2018}, while in IS the contribution from the disk and/or from the NS increases, cools down the corona and reduces the X-ray variability. The distinction between sources in IS and EIS is not strict, and some sources, like 4U 0614+091 itself, have been found in both states \citep[see panel 1 in Figure 2 of ][]{VanStraaten2002}. 
Assuming that radio jet spectra in NS LMXBs are usually flat\footnote{Aside of 4U 0614+091, a couple of other examples can be found in \cite{DiazTrigo2017}}, it is plausible that X-ray PDS can be used as proxy of the variability in the ejecta for NS LMXBs in EIS. In this sense, AMXPs and/or low luminosity bursters, usually found in EIS, could be good candidates to test ISHEM in the future. \\
\section{Conclusions}
In this work we presented the first ever attempt to describe the broadband emission of a NS LMXB, i.e. 4U 0614+091, with a model taking into account both the jet and the accretion flow emission. We took advantage of the same multi-wavelength data set presented by \cite{Migliari2010}, with the only exception for the \emph{Swift}/UVOT data, which were re-analyzed. We modelled the radio-to-IR spectrum with the ISHEM code, which calculates the expected spectral energy distribution in the low energy part of the SED taking into account the X-ray variability (connected in turn to the internal shocks temporal pattern). In particular, we used the quasi-simultaneous \emph{Swift}/XRT PDS as input for the "synthetic" SED. While the ISHEM model has been applied several times in the past to X-ray binaries hosting BHs as the accreting object, this is the first time that the model is applied to a system hosting a NS. In addition, optical-to-X-ray data were modelled with an irradiated disc model. \\
We found that the compatibility between the SED built using the X-ray PDS and the data set is critically dependent on the geometry of the jet, enclosed in the geometrical parameter $\zeta$. In particular, a highly non-conical geometry, with $\zeta\approx 0.6$, results in an acceptable fit. Alternatively, an acceptable fit is found within a conical geometry scenario but using in input a "flicker-noise" PDS instead of the X-ray PDS. This scenario might imply that for NS LMXBs the X-ray PDS are not good tracers for the fluctuations in the Lorentz factors of the ejecta, possibly due to some contribution from the boundary layer/NS emission. The scarce statistics does not allow for the moment to choose one scenario over the other. New observations and/or further studies like the one presented here are definitely necessary to provide an answer to this issue and in general for a better understanding of the accretion-ejection coupling in NS LXMBs.

\section*{Acknowledgements}
This work received financial support from PNHE in  France  and  from  the  OCEVU  Labex  (ANR-11-LABX-0060) and the A*MIDEX project (ANR-11-IDEX-0001-02) funded by the  ‘Investissement d’Avenir’ French government program  managed  by  the  ANR.
We acknowledge financial contribution from the agreement ASI-INAF n.2017-14-H.0 and INAF main-stream (P.I. Belloni). JLM acknowledges the support of a fellowship from "La Caixa" Foundation (ID 100010434). The fellowship code is LCF/BQ/DR19/11740030. MP acknowledges financial support from the Spanish Ministry of Science through Grants PID2019-105510GB-C31, PID2019-107427GB-C33 and AYA2016-77237-C3-3-P, and from the Generalitat Valenciana through grant PROMETEU/2019/071.

\bibliography{biblio}
\newpage
\appendix
\section{The updated version of ISHEM} \label{sec:model}

\begin{figure}
\centering
\includegraphics[scale=0.35]{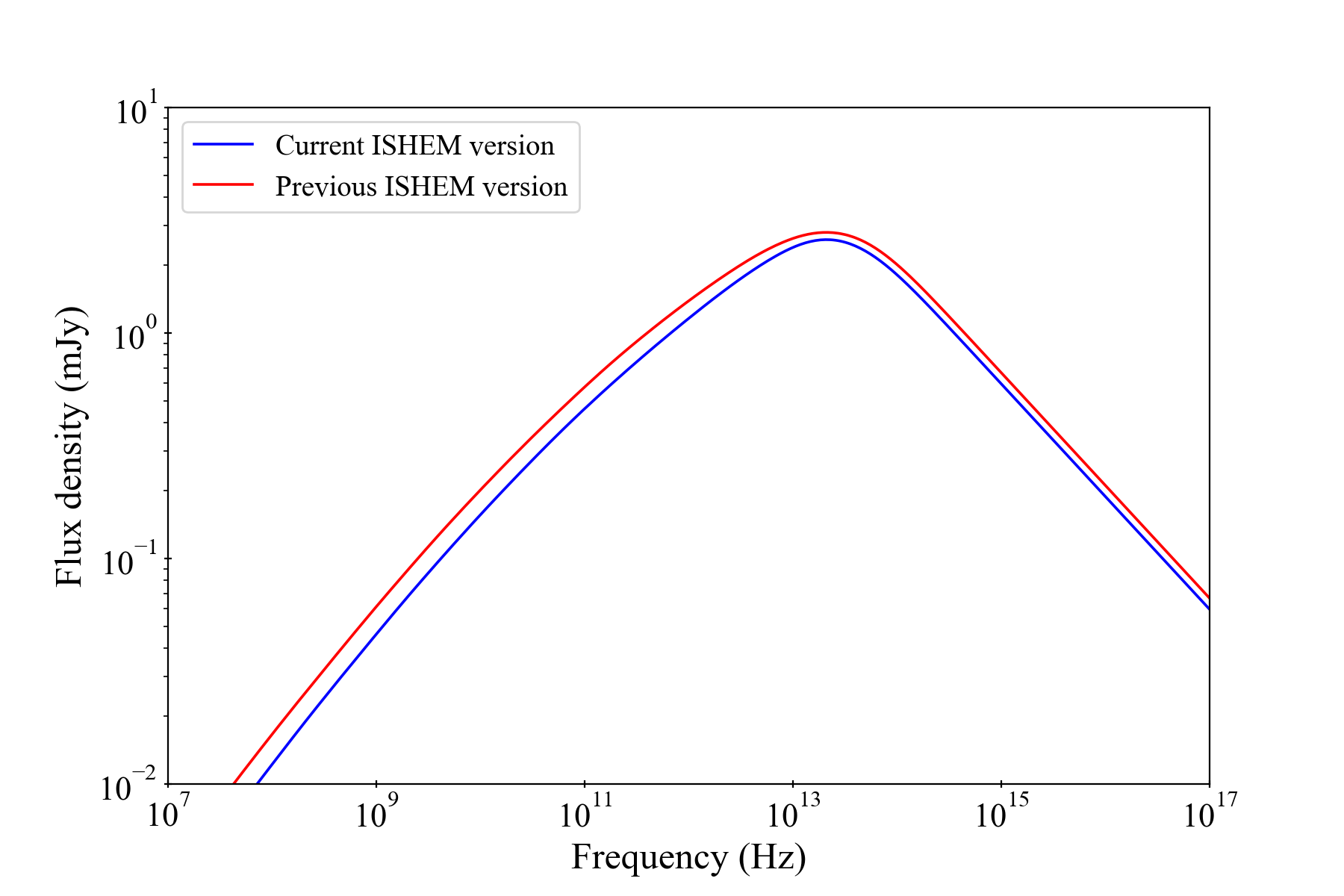}
\caption{Comparison between the SED simulated with the previous (red) and the updated (blue) version of ISHEM. In both simulations, $\zeta=1.0$ and $t_{\rm simu}=10^5$ s.}
\label{fig:versions}
\end{figure}

The {\sc ishem} code is extensively detailed in \citet{Malzac2014}.
However in this paper we use an updated version of {\sc ishem} in which radiation transfer was improved in order to account more accurately for the geometrical and relativistic aberration effects  on the synchrotron process. The main effects of these modifications is to change the normalisation of the predicted SED by at most a factor of a few compared to the previous version. The shape of the predicted SED is not significantly affected (see Fig. \ref{fig:versions}). Although we expect that the new version is more accurate and provides better estimates of the jet parameters when compared to the data, from a qualitative point of view, the resulting parameters are comparable to those obtained with the previous version. The main changes in the code are described below.

\subsection{Emission and absorption coefficients}

In the version of {\sc ishem} presented in \citet{Malzac2014}, we used the synchrotron emission and absorption coefficient given in \citet{Chaty11}. These estimates are for a uniform magnetic field observed with a specific line of sight that is perpendicular to the magnetic field \citep{RL86}. In the new version of the code, it is instead assumed the field is tangled on scales larger than the Larmor radius and smaller than the emitting region. This constitutes a better approximation of the magnetic field in a shocked region and the result is valid for any viewing angle \citep{Crusius86}.
We detail below the formulae that we used for the synchrotron emissivity and absorption coefficients in a tangled magnetic field.

For a given emitting electron of Lorentz factor $\gamma$, emitting or absorbing photons at a frequency $\nu$,  we define the reduced photon frequency: 
\begin{equation}
x=\frac{\nu}{C_1 B \gamma^2},
\end{equation}
where $B$ the amplitude of the magnetic field, $\nu$ the emitted photon frequency and,
\begin{equation}
    C_1=\frac{3 q}{4\pi m c},
\end{equation}
where $m$ is the electron rest mass, $c$ the speed of light $q$ the electric charge of the electron.

The pitch angle averaged emissivity (erg/s/ster/cm$^{3}$), for a particle energy density distribution $N(\gamma)$ (in cm$^{-3}$) can be written as:

\begin{equation}
j_\nu=C_2 B \int_0^{+\infty}N(\gamma) R(x) d\gamma
\label{eq:jnu}
\end{equation}
where
\begin{equation}
C_2=\sqrt{3}q^{3}/(4\pi m c^2)\simeq 1.8655582 \times 10^{-23}  \quad\mathrm{cgs}
\end{equation}
and
\begin{equation}
R(x)=\frac{x^2}{2}K_{\frac{4}{3}}\left(\frac{x}{2}\right)K_{\frac{1}{3}}\left(\frac{x}{2}\right)-\frac{3x^3}{20}\left[K_{\frac{4}{3}}^2\left(\frac{x}{2}\right)-K_{\frac{1}{3}}^2\left(\frac{x}{2}\right)\right]
\end{equation}
where $K_m$ is the modified Bessel function of order $m$ \citep{Ghisellini88}. This formula is equivalent to the one given by \citet{Crusius86} in terms of Whitaker functions. 

For a power law particle energy distribution,
\begin{equation}
N(\gamma)=N_0\gamma^{-p},
\label{eq:ng}
\end{equation}
Equation~\ref{eq:jnu} can be integrated analytically \citep{Crusius86}. This gives:
\begin{equation}
j_\nu(\nu)=G(p)C_2 B N_0\left(\frac{\nu}{C_1 B}\right)^{\frac{1-p}{2}}
\label{eq:powjnu}
\end{equation}

with

\begin{equation}
G(p)=\sqrt{\frac{\pi}{2^{7-p}}}\, \frac{p+7/3}{p+1}\,\Gamma\left(\frac{3p-1}{12}\right)\Gamma{\left(\frac{3p+7}{12}\right)}\frac{\Gamma\left(\frac{p+5}{4}\right)}{\Gamma\left(\frac{p+7}{4}\right)}
\end{equation}
 of order of  unity: $G(2)\simeq0.7485$, $G(3)\simeq0.5374$, and $\Gamma$ represents the usual gamma function.

The absorption coefficient (cm$^{-1}$) is \citep{Ghisellini91}:
\begin{equation}
\alpha_\nu=-\frac{C_2 B}{2m\nu^2}\int_0^{+\infty}R(x)\gamma^2\frac{d N(\gamma)\gamma^{-2}}{d\gamma} d\gamma 
\end{equation}

For the power-law distribution given by Equation~\ref{eq:ng} we have to calculate the same integral as in Equation~\ref{eq:jnu} but with an electron index replaced is  $p+1$ instead of $p$. This gives:
\begin{equation}
\alpha_\nu=\frac{(p+2) G(p+1) }{2} \frac{C_2 B N_0}{m\nu^2} \left(\frac{\nu}{C_1 B}\right)^{-p/2}
\label{eq:alphanu}
\end{equation}

Then the source function is simply:
\begin{equation}
S_\nu=\frac{j_\nu}{\alpha_\nu}=\frac{2G(p)m \left(C_1 B\right)^{-\frac{1}{2}}}{(p+2) \, G(p+1)} \nu^{\frac{5}{2}}
\label{eq:snu}
\end{equation}

These expressions for emission and absorption coefficients  are equivalent to those used in a different form by \citet{Zdziarski12}.

\subsection{Emission from a homogeneous cylinder in motion}
In {\sc ishem} the jet is discretized into a large number of homogeneous cylinders whose axes are along the jet axis. These cylindric shells travel along the jet while expanding radially according to the fixed jet geometry. The time-dependent emission from each of these cylinders is calculated by {\sc ishem}. In order to predict the time-averaged SED of the jet, it is necessary to calculate the time-integrated emission of millions of such cylinders. For reasons of computational efficiency, the radiation transfer has to be simplified. 

In the original version of the code the instantaneous flux received by the observer was simply:
\begin{equation}
    F_{\nu}= \delta^3 \frac{\tilde{H}R}{2 D^2} \tilde{S}_{\tilde{\nu}}\left[1-\exp\left(-\tilde{\alpha}_{\tilde{\nu}} R\right)\right] 
    \label{eq:fnum14}
\end{equation}
where $\delta$ is the standard relativistic Doppler factor of the cylinder, $\tilde{H}$ is the height of the cylinder, $D$ the distance of the source  and $R$ the radius of the cylinder (in this section tilded symbols represent quantities measured in the rest frame of the cylinder). The estimate given by equation~\ref{eq:fnum14} does not take into account the changes in the projected surface area of the cylinder when observed at different angle and it does not account for the relativistic aberration effects. Also, the emission of each shell is calculated independently, the possible effects of absorption by other shells along the line of sight were neglected. We propose below an improved treatment of the geometrical effects which is implemented in the new version of the code. 

\subsubsection{In the rest frame}

For now, let us consider the emission of a cylinder in its rest frame, and neglect the absorption of radiation by the other parts of the jet. 
The power spectral density per unit solid-angle emitted in a direction $\vec{n}$ making angle $\tilde{\theta}$ with respect to the velocity of the cylinder 
\begin{equation}
\frac{d\tilde{L}_(\tilde{\mu})}{d\tilde{\Omega}d\tilde{\nu}}=\int_{\tilde{A}_\perp(\tilde{\mu})} \tilde{S}_{\tilde\nu}\left[1-\exp\left(-\tilde{\alpha}_{\tilde{\nu}} \tilde{l}\right)\right] d\tilde{A}
\end{equation}
where the integration is over the area $\tilde{A}_{\perp}$ of the projection of the cylinder on to $P$, a plane normal to $\vec{n}$:
\begin{equation} 
\tilde{A}_\perp= 2R\tilde{H} \sin{\tilde{\theta}}+\pi R^2 \arrowvert \tilde{\mu}\arrowvert
\end{equation}
where $\tilde{\mu}=\cos{\tilde{\theta}}$.
This formula works also if the cylinder is seen from below (i.e. with $\tilde{\mu} <0$).

$\tilde{l}$ measures the local physical depth of the cylinder accross the direction $\vec{n}$, $\tilde{l}$ is a complicated function of the position on  the integration plane P and $\tilde{\mu}$.
In order to simplify the calculation, we replace the complicated function $\tilde{l}$ by the average depth of the cylinder across the direction $\vec{n}$:
\begin{equation}
\langle\tilde{l}\rangle=\frac{\tilde{V}}{\tilde{A}_\perp}=\left[ \frac{2\sin{\tilde{\theta}}}{\pi R}+\frac{\arrowvert \tilde{\mu}\arrowvert }{\tilde{H}}\right]^{-1}.
\label{eq:lav}
\end{equation}
With these approximations, the emission pattern of the source is given by:
\begin{equation}
\frac{d\tilde{L}_(\tilde{\mu})}{d\tilde{\Omega}d\tilde{\nu}}=\tilde{A}_\perp \tilde{S}_{\tilde\nu}\left[1-\exp\left(-\tilde{\alpha}_{\tilde{\nu}} \langle\tilde{l}\rangle\right)\right].
\end{equation}
We note that this expression becomes exact in both the optically thin and thick limits.

\subsubsection{In the observer's frame}

In the observer's frame:
\begin{equation}
\frac{dL_(\mu) }{d\Omega d\nu}=\delta^3 \frac{d\tilde{L}_(\tilde{\mu})}{d\tilde{\Omega}d\tilde{\nu}}.
\end{equation}
And the relation between $\mu$ and $\tilde{\mu}$ is given by:
\begin{equation}
\tilde{\mu}=\frac{\mu-\beta}{1-\mu\beta},
\end{equation},
\begin{equation}
\sin{\tilde{\theta}}=\delta \sin{\theta},
\end{equation}
and of course:
\begin{equation}
\nu=\delta\tilde{\nu}.
\end{equation}
(cf \cite{RL86})

The observed flux at a large distance $D$ in the direction $\mu$ is then:
\begin{equation}
F(\nu)=\frac{dL_(\mu) }{d\Omega d\nu}D^{-2}=\delta^3 \frac{\tilde{A}_\perp}{D^2} \tilde{S}_{\tilde{\nu}}\left(\frac{\nu}{\delta}\right) \left[1-\exp\left(-\tilde{\alpha}_{\tilde{\nu}}\left(\frac{\nu}{\delta}\right) \langle\tilde{l}\rangle\right)\right],
\label{eq:fnus}
\end{equation}
with
\begin{equation} 
\tilde{A}_\perp= \delta \left[2R\tilde{H} \sin{\theta} +\pi R^2 \gamma \arrowvert{\mu-\beta}\arrowvert\right]
\end{equation}
and
\begin{equation}
\langle\tilde{l}\rangle=\delta^{-1} \left[ \frac{2\sin{\theta}}{\pi R}+\frac{\gamma\arrowvert\mu-\beta\arrowvert}{\tilde{H}}\right]^{-1}
\label{eq:lav0}
\end{equation}
Using the standards transformations \citep[see, e.g., ][]{Ghisellini00}:
\begin{equation}
{S_\nu}=\delta^3\tilde{S}_{\tilde{\nu}}, \quad {\alpha_\nu}=\tilde{\alpha}_{\tilde{\nu}}/\delta, \quad V=\delta \tilde{V},\quad \langle l\rangle=\delta \langle\tilde{l}\rangle,
\end{equation}
we can recover  the expression of the received flux in terms of quantities measured in the observer's frame:

\begin{equation}
F(\nu)=\frac{{A}_\perp}{D^2} {S}_{{\nu}} \left[1-\exp\left(-{\alpha}_{{\nu}} \langle{l}\rangle\right)\right]
\label{eq:fnufinal}
\end{equation}

Note that Equation~\ref{eq:fnufinal} implies that the projected area is the same in both frames, as expected:
\begin{equation}
A_\perp=\tilde{A}_\perp.
\end{equation}
Indeed, the net relativistic effect is to rotate the apparent viewing angle of the cylinder by an angle $\cos{\alpha}=\beta$ in the observer's frame. But it looks exactly as in the rest frame, there is no contraction of the image of the cylinder (see \cite{Ghisellini00}).

\subsubsection{Effects of absorption by other shells}

The radiation escaping through the top of the shell is absorbed by the others shells in the jet located between the shell surface and the observer, while on the other hand  the lateral  section of the cylinder is observed directly and is not affected.
To take this into account we use an effective synchrotron absorption depth $\tau_{e}$ along the line of sight to reduce accordingly the emission escaping through the top (or bottom surface of a shell) by e factor $e^{-\tau_{e}}$.
This is equivalent to replacing  $A_\perp$ in equation~\ref{eq:fnufinal} by an effective projected surface area:

\begin{equation}
A_{\perp,e}=\delta \left[2R\tilde{H} \sin{\theta} +\pi R^2 \gamma \arrowvert{\mu-\beta}\arrowvert e^{-\tau_e}\right]
\label{eq:aperpabs}
\end{equation}

To simplify the calculations we consider absorption by other parts of the jet in a time averaged sense only. We use the simulation to estimate a time-averaged absorption coefficient $\langle\tilde{\alpha}_{\nu}\rangle(z)$ in the jet, where $z$ is the distance from the base of the jet. 
From this we can tabulate a function $\tau_j$:

\begin{equation}
\tau_j=\int_z^{+\infty}\frac{ \langle\tilde{\alpha}_{\nu}\rangle}{\delta_j\mu} dz
\end{equation}
To simplify the calculation we consider an average line of sight which goes through the centre of the cylinder which is located at an instantaneous  position $z_c$. We then calculate the position $z_i$ at which a light ray traveling along this line of sight escapes from the jet towards the observer. We then estimate $\tau_e$ as:
\begin{equation}
\tau_e=\tau_j(z_s)-\tau_j(z_i)
\end{equation}
for $\mu >0$
and 
\begin{equation}
\tau_e=\tau_j(z_i)-\tau_j(z_s)
\end{equation}
for $\mu<0$.

\section{Scaling laws for parabolic jets}\label{sec:scaling}

In this section we derive the scaling laws for the synchrotron spectral break frequency and flux that we used to estimate the best fits {\sc ISHEM} parameters shown in Figs.~\ref{fig:areas} and~\ref{fig:angles}.

\subsection{Synchrotron emission of self-similar parabolic jets}

First, we are have to estimate the SED of standard self-similar parabolic jets, i.e. jets with radius increasing with height like $R=R_0\left(z/z_0\right)^{\zeta}$, where $z_0$ and $R_0$ are the height and radius at the base of the jet emitting region. 

In this section we do not presume anything about the jet dissipation mechanism, i.e. it does not have to be necessarily dominated by internal shocks. However we assume that, as in the internal shock model, the jet is made of homogeneous cylindric shells of proper vertical scale $\tilde{H}$ and radius following the parabolic dependence with $z$ defined above. Each of these shells emits an instantaneous observed flux $F_j$. The total jet flux is the time averaged flux of a shell as it crosses the whole jet multiplied by the total number of shells $n_{\rm s}$ present at any time in the jet:
\begin{equation}
F_j=  \frac{n_s}{t_{rf}-t_{r0}} \int_{t_{r0}}^{t_{rf}} F dt_r, 
\end{equation}
where $t_{rf}-t_{r0}= (z_f-z_0)(1-\beta\mu)/\beta /c$ is the shell jet crossing time as measured by the observer (assuming that the jet emitting region starts at $z_0$ and ends at $z_f$). The number of shells is given by the jet size divided by the observed length of the shells and corrected by their volume filling factor $f_v$:  $n_s=(z_f-z_0) f_v/\delta \tilde{H}$. The total jet flux can be rewritten as: 
\begin{equation}
F_j= \int_{z_0}^{z_f} \frac{f_v F}{\delta \tilde{H}} dz, 
\label{eq: flint}
\end{equation}
where F is given by equation~\ref{eq:fnus}.

Let us assume that the magnetic field in the emitting shells decreases with the jet radius like $R$:
\begin{equation}
B(R)=B_0 (R/R_0)^{-b}
\end{equation}
We assume that througout the jet, the electron energy distribution inside the shells is a power-law of index $p$ as given by equation~\ref{eq:ng}, within the range of electron Lorentz factors $\gamma_{\rm min}$--$\gamma_{\rm max}$ with $\gamma_{\rm max} >> \gamma_{\rm min}$.
We assume a constant ratio $\xi_e$ between the particle kinetic and magnetic energy densities so that:
\begin{equation}
N_0=\frac{\xi_e i_\gamma  B^2}{8\pi mc^2},    
\end{equation}
where 
\begin{equation}
i_\gamma=(j_\gamma-k_\gamma)^{-1}.
\end{equation}
For $p=1$:
\begin{equation}
k_\gamma=\ln(\gamma_{\rm max}/\gamma_{\rm min}),
\end{equation}
otherwise:
\begin{equation}
k_\gamma=\frac{\gamma_{\rm max}^{1-p}-\gamma_{\rm min}^{1-p}}{1-p}.
\end{equation}
For p=2:
\begin{equation}
j_\gamma=\ln(\gamma_{\rm max}/\gamma_{\rm min}),
\end{equation}
otherwise:
\begin{equation}
j_\gamma=\frac{\gamma_{\rm max}^{2-p}-\gamma_{\rm min}^{2-p}}{2-p}.
\end{equation}

\noindent Under these assumptions, the synchrotron absorption coefficient is well approximated by equation~\ref{eq:alphanu}. It can be rewritten as:
\begin{equation}
\tilde{\alpha}_{\tilde{\nu}}=\tilde{\alpha}_0 \left(R/R_0\right)^{-bd}=\tilde{\alpha}_0 \left(z/z_0\right)^{-\zeta bd},
\end{equation}
with
\begin{equation}
\tilde{\alpha}_0=K_\alpha \xi_e i_\gamma B_0^d \left(\frac{\nu}{\delta}\right)^{-\frac{p+4}{2}},
\end{equation} 
and $d=3+p/2$. 
The constant $K_\alpha$ is given by:
\begin{equation}
K_{\alpha}=\frac{(p+2) G(p+1) }{2} \frac{C_1^{p/2}C_2}{8\pi m^2c^2} 
\end{equation}

\noindent The source function given by equation~\ref{eq:snu} becomes:
\begin{equation}
\tilde{S}_\nu=S_x \left(z/z_0\right)^{\zeta b/2}
\end{equation}
with
\begin{equation}
S_x= \frac{K_j}{K_\alpha} B_0^{-1/2}\left(\frac{\nu}{\delta}\right)^{5/2}
\end{equation}
and
\begin{equation}
\frac{K_j}{K_\alpha}=\frac{2G(p)m C_1^{-\frac{1}{2}}}{(p+2) \, G(p+1)}.
\end{equation}

In the following, we assume that $\zeta bd >1$  and $z_f>>z_0$.

\subsection{Local cylinder approximation}\label{localcylapp}

If we choose $\tilde{H}>>R$, the projected area and photons crossing length (equations \ref{eq:aperpabs} and \ref{eq:lav0}) become:
\begin{equation} 
\tilde{A}_\perp\simeq 2\tilde{H} R \delta\sin{\theta}
\label{eq:apeprplargeangles}
\end{equation}
\begin{equation}
\langle\tilde{l}\rangle\simeq\frac{\pi R}{2\delta \sin{\theta}}
\label{eq:lavlargeangles}
\end{equation}
This corresponds to the local cylinder approximation of the jet. In this approximation the variations of the source function and absorption coefficients along the line of sight are neglected.

This approximation is expected to be accurate at large inclination angles ($\theta \sim 90^\circ$) where the observed emission is dominated by radiation travelling in the radial direction and does not experience significant gradients in the jet. In fact it turns out to be remarkably accurate even at smaller viewing angles and up to viewing angles comparable to the jet opening angle (i.e. up to $\tan\theta\sim R_0/Z_0)$. For smaller jet inclinations, a different approach must be adopted \citep[see, e.g.][]{Zdziarski2016}

In the framework of the local cylinder approximation, the jet flux can be rewritten as:
\begin{equation}
F_j=\frac{\delta^3\sin{\theta}}{\zeta(bd-1)}\frac{2f_v R_0z_0 S_x}{D^2 \tau_1^{a_1-1}} F_1(a_1-1,\tau_f,\tau_1)
\end{equation}
where:
\begin{equation}
\tau_1=\frac{\alpha_0\pi R_0}{2\delta\sin{\theta}}
\end{equation}

\begin{equation}\label{eq:tauf1}
\tau_{f1}=\tau_1(z_f/z_0)^{\zeta(1-bd)}
\end{equation}
\begin{equation}
a_1=1+\frac{2+2\zeta+\zeta b}{2\zeta(1-bd)}
\end{equation}
\begin{equation}
F_1(x,y,z)=\int_{y}^{z} \tau^{x-1} (1-e^{-\tau}) d\tau
\end{equation}

The jet emits in the optically thin regime at frequencies for which $\tau_f<<1$ and $\tau_1<<1$. In this case $F(a_1-1,\tau_f,\tau_1)\simeq \chi_{1} \tau_1^{a_1}$ with,
\begin{equation}
\chi_{1}=\frac{1-x_f^{\zeta(1-bd)a_1}}{a_1}.
\end{equation}
The optically thin flux can be expressed as:
\begin{equation}
F_{j,thin}=\frac{\delta^3\sin{\theta}}{\zeta(bd-1)}\frac{2f_v \chi_1 R_0z_0 S_x\tau_1}{D^2} 
\end{equation}
On the other hand, the partially absorbed emission corresponds to frequencies at which $\tau_f<<1$ and $\tau_1>>1$. Then for $b>0$, $F_1(a_1-1,\tau_{f1},\tau_1)\simeq -\Gamma(a_1-1)$, and the partially absorbed flux can be expressed as:
\begin{equation}
F_{j,abs}=\frac{-\delta^3\sin{\theta}}{\zeta(bd-1)}\frac{2f_v R_0z_0 S_x}{D^2 \tau_1^{a_1-1}} \Gamma(a_1-1)
\end{equation}
The break frequency $\nu_b$ is defined as the transition between these two regimes. It occurs at the frequency $\nu_b$ at which $F_{j,thin}=F_{j,abs}$:
\begin{equation}
\nu_b=\delta\left(\frac{2\delta\sin{\theta}}{\pi R_0B_0^dK_\alpha \xi_e}\right)^{-\frac{2}{p+4}}\left[\frac{-\Gamma(a_1-1)}{\chi_1}\right]^{-\frac{2}{(p+4)a_1}}
\label{eq:nubpar}
\end{equation}
The flux at the break frequency is then given by:
\begin{equation}
F_{\nu_b}=\delta^2\left(\delta \sin{\theta}\right)^\frac{p-1}{p+4} \frac{2f_v K_j}{K_{\alpha}}\frac{ \left(\pi K_\alpha \xi_e/2\right)^{\frac{5}{p+4}}}{\zeta(bd-1)} 
\frac{\chi_1 z_0 R_0^{\frac{p+9}{p+4}} B_0^\frac{2p+13}{p+4}}{\left(-\Gamma(a_1-1)/\chi_1\right)^\frac{1-p}{(p+4)a_1}}
\label{eq:fnubpar}
\end{equation}


\subsection{Scaling laws in the internal shock model}

The parameters of the base of the jet emitting region $z_0$,$R_0$ and $B_0$, depend on the model for the dissipation in the jet. In the case of internal shocks \citep{Malzac2013,Malzac2014}:
\begin{equation}
z_0\propto \Gamma_{\rm av}\beta(\Gamma_{\rm av}+1)
\label{eq:z0}
\end{equation}
\begin{equation}
B_0\propto P_j^{1/2} \left[\Gamma_{\rm av}\beta(\Gamma_{\rm av}+1)\right]^{-\frac{1+2\zeta}{2}}  \frac {z_b^\zeta}{R_b}
\label{eq:b0}
\end{equation}
and
\begin{equation}
R_0 \propto \left(\frac{z_b^\zeta}{R_b}\right)^{-1} \left[ \gamma\beta(\gamma+1)\right]^\zeta
\label{eq:r0}
\end{equation}
where $R_b$ is the radius of the jet at the point of ejection (i.e; close to the compact object). This point is located at height $z_b=z_0 (R_b/R_0)^{1/\zeta}$. We defines the jet opening angle $\phi$ such that $\tan{\phi}=R_{b}/z_{b}$ i.e. as in conical geometry although in parabolic geometry with $\zeta < 1$, the actual $R/z < \tan{\phi}$ at $z> z_b$.
Note that the scaling relations given by equations~\ref{eq:z0},\ref{eq:b0} and \ref{eq:r0}  are obtained in the limit of $z_0 >> z_b$. They may not work very well at large $z_b$ i.e. very small jet opening angles.

Injecting them into equations~\ref{eq:nubpar} and \ref{eq:fnubpar} we obtain finally the dependence of $\nu_b$ and $F_{\nu_b}$ on the main parameters of {\sc ishem}:

\begin{equation}
\nu_b\propto    \delta\left(\delta\sin{\theta}\right)^\frac{-2}{p+4}  \left(i_\gamma \xi_e \right)^{\frac{2}{p+4}} \frac{z_b^\zeta}{R_b}  \left[\Gamma_{\rm av}\beta(\Gamma{\rm av}+1)\right]^{-\frac{6+8\zeta+(1+2\zeta)p}{2p+8}} P_j^\frac{p+6}{2p+8}
\end{equation}
\begin{equation}
F_{\nu_b} \propto \frac{\delta ^2}{D^2}\left(\delta\sin{\theta}\right)^{\frac{p-1}{p+4}} \left(\xi_e i_\gamma\right)^\frac{5}{p+4}  \frac{z_b^\zeta}{R_b}\left[\Gamma_{\rm av}\beta(\Gamma_{\rm av}+1)\right]^{-\frac{5+8\zeta+2\zeta p}{2p+8}} P_j^\frac{2p+13}{2p+8}
\end{equation}

These scaling laws are expected to constitute a good approximation for inclinations $\theta > \phi_0$ where $\phi_0$ is the jet opening angle at the base of the jet emitting region $\tan\phi_0=R_0/Z_0$. Since in the case of 4U 0614+091 the first internal shocks occur  around $z_0\sim 10^3  R_g$, and in all our models we have set $R_b=$10 $R_{\rm G}$ the approximation is expected to be valid for inclinations such that $\tan{\theta} > 10
^{-2(1-\zeta)} \tan{\phi}^{\zeta}$, where $\phi$ is the jet  opening angle at $R_b$. As they are independent of the magnetic field profile $b$, the scaling relations are valid for a wide range of dissipation profile along the jet. They extend the formulae used in \cite{Peault2019} to non-conical geometries. The angle dependence of the scaling relations are also improved  with respect to the formulae of \cite{Peault2019}) in order to reflect the refined treatment of the anisotropy of the jet radiation implemented in the new version of {\sc ishem}.  
Note however that they do not take into account the possible contribution of the counter jet. 

\label{lastpage}
\end{document}